\documentstyle[12pt]{article}

\begin{document}
\makeatletter \@addtoreset{equation}{section}
\renewcommand{\theequation}{\thesection.\arabic{equation}}

\baselineskip=18.6pt plus 0.2pt minus 0.1pt


\begin{titlepage}
\title{
\hfill\parbox{4cm} {\normalsize UFR-HEP/02-11}\\ \vspace{1cm}
     {\bf   NC  Calabi-Yau    Orbifolds   in Toric Varieties with Discrete Torsion  }
}
\author{
A. Belhaj\thanks{{\tt ufrhep@fsr.ac.ma}},   and E.H.
Saidi\thanks{{\tt H-saidi@fsr.ac.ma}} {}
\\[7pt]
{\it Lab/UFR-High Energy Physics, Faculty of Sciences, Rabat,
Morocco.} }

\maketitle \thispagestyle{empty}
\begin{abstract}
  Using the algebraic geometric approach of Berenstein et {\it al}
(hep-th/005087 and hep-th/009209) and methods of toric geometry,
we study non commutative (NC) orbifolds of Calabi-Yau
hypersurfaces   in toric varieties   with discrete torsion. We
first develop a new  way of getting  complex $d$ mirror Calabi-Yau
hypersurfaces $H_{\Delta }^{\ast d}$   in toric manifolds
$M_{\Delta }^{\ast (d+1)}$ with a $C^{\ast r}$ action and analyze
the general group of the discrete isometries of $H_{\Delta }^{\ast
d}$.  Then we build a general class of $d$ complex dimension NC
mirror Calabi-Yau orbifolds where the  non commutativity
parameters $\theta _{\mu \nu }$ are solved in terms of discrete
torsion and toric geometry  data of $M_{\Delta }^{(d+1)}$ in which
the original  Calabi-Yau hypersurfaces is embedded. Next we work
out a generalization of the NC algebra for generic $d$ dimensions
NC Calabi-Yau manifolds and  give various representations
depending on different choices of the Calabi-Yau toric  geometry
data. We also study
fractional D-branes at orbifold points. We   refine  and extend the result for   NC $%
(T^{2}\times T^{2}\times T^{2})/(\mathbf{{Z_{2}}\times {Z_{2})}}$
 to higher dimensional torii orbifolds in terms of Clifford algebra.
\end{abstract}

\newpage
\end{titlepage}
\newpage
\def\be{\begin{equation}}
\def\ee{\end{equation}}
\def\bea{\begin{eqnarray}}
\def\eea{\end{eqnarray}}
\def\nn{\nonumber}
\def\l{\lambda}
\def\t{\times}
\def\[{\bigl[}
\def\]{\bigr]}
\def\({\bigl(}
\def\){\bigr)}
\def\p{\partial}
\def\o{\over}
\def\ta{\tau}
\def\cm{\cal M}
\def\R{\bf R}
\def\b{\beta}
\def\a{\alpha}
\tableofcontents
\newpage

\section{Introduction}
Non-commutative (NC) geometry plays an interesting role in the
context of string theory \cite{1} and  in  compactification of the
Matrix model  formulation of M-theory on NC torii \cite{2-7},
which has opened new lines of research devoted to the study  NC
quantum field theories \cite{8};  see  also [9-23]. In the context
of string theory, NC geometry is involved whenever an
antisymmetric $B$-field
 is turned on. For example,  in the study of  the ADHM construction $D(p-4)/Dp$ brane
 systems ($p>3$) \cite{24},
   the NC version of the Nahm construction for monopoles \cite{25}
   and
   in the study of
   tachyon condensation using the so called GMS approach \cite{26},
    see also[27-31].
\par More recently efforts
have been devoted to go beyond the particular NC $R_{\theta
}^{d}$, NC T$_{\theta }^{d}$   geometries [29-37]. A special
interest has been given to build NC Calabi-Yau manifolds
containing the commutative ones as subalgebras and a development
has been obtained for the case of orbifolds of Calabi-Yau
hypersurfaces. The key point of this construction, using a NC
algebraic geometric method \cite{38}, see also \cite{39,40}, is
based on solving non commutativity in terms of discrete torsion of
the orbifolds.  In this regards, there  are two ways one may
follow to construct this extended geometry. (i) A constrained
approach using purely geometric  analysis, which we are interested
in this paper. (ii) Crossed product algebra based on the
techniques of the fibre bundle  and the   discrete group
representations.  For the first way, it has been
shown that the $\frac{T^{2}\times T^{2}\times T^{2}}{\mathbf{{Z_{2}}\times {%
Z_{2}}}}$, orbifold of the product of three elliptic curves with
torsion, embedded in the $C^{6}$ complex space, defines a NC
Calabi-Yau threefolds \cite{39} having  a remarkable
interpretation in terms of string states. Moreover on the fixed
planes of this NC threefolds, branes fractionate and local
 complex deformations are no more trivial. This  constrained  method  was also  applied
  successfully   to
 Calabi-Yau hypersurfaces described by homogeneous
polynomials with discrete symmetries including  $K3$ and the
quintic  as particular geometries \cite{39,40,41,42,43}.    NC
algebraic geometric approach for building NC Calabi-Yau manifolds
has very remarkable features and  is suspected to have deep
connections both with the intrinsic properties of toric varieties
\cite{44,45,46} and the R matrix of Yang-Baxter equations of
quantum spaces \cite{47,48,49}.
\par In this study we  extend the Berenstein and Leigh ( $BL$ for short) construction for
 NC Calabi-Yau manifolds with discrete
torsion by considering $d$ dimensional complex Calabi-Yau
orbifolds embedded in  $(d+1)$ complex toric manifolds  and using
toric geometry method \cite{50,51,52,53}.    In particular,   we
build a general class of $d$ complex dimension non commutative
mirror Calabi-Yau orbifolds  for which  the  non commutativity
parameters $\theta _{\mu \nu }$ are solved in terms of discrete
torsion and toric geometry data of dual polytopes ${\Delta }
(M^d)$.  To establish these  results,  we will proceed   in three
steps:\\ $\mathit{(i)}$ We consider   pairs of mirror Calabi-Yau
hypersurfaces $H_{\Delta }^{d}$ and $H_{\Delta }^{\ast d}$\
respectively embedded in the toric manifolds $M_{\Delta }^{d+1}$ and $%
M_{\Delta }^{\ast (d+1)},$ where $\Delta $\ is their attached
polyhedron;
and develop a manner of handling these spaces by working out the explicit solution for
 the so called $%
Y_{\alpha }=\Pi _{i=1}^{k+1}$\ $x_{i}^{<{V}_{i},{V}_{\alpha
}^{\ast }>}$\ invariants of the $C^{\ast r}$ actions  and their mirrors $%
y_{i}=\Pi _{I=1}^{k^{\ast }}$\ $z_{I}^{<{V}_{I}^{\ast },{V}%
_{i}>}$.  The construction we will give  here is a new one; it is
based on pushing further the solving of the Calabi-Yau constraint
eqs regarding the invariants under the $C^{\ast r}$ toric actions.
Aspects of this analysis may be approached with the analysis of
\cite{53,54}, but  the  novelty is in the
manner we treat the $C^{\ast r}$ invariants. Then we fix our attention on $%
H_{\Delta }^{\ast d}$ described by the zero of a homogeneous polynomial $%
P_{\Delta }(z)$\ of degree $D$ and explore the general form of the
group of discrete symmetries $\Gamma $ of $H_{\Delta }^{\ast d}$\
using the toric  geometry data $\{q_{i}^{a};{V}_{i}\ ;1\leq i\leq
k+1$;$\quad 1\leq a\leq r;\quad d=(k-r)\}$ of the polyedron
$\Delta $.\\ $\mathit{(ii)}$  We show that for  the  special
region in the moduli space  where complex deformations are set to zero, the polynomials $%
P_{\Delta }$ defining the Calabi-Yau hypersurfaces have a larger
group of discrete symmetries $\Gamma _{0}$ containing as a
subgroup the usual $\Gamma _{cd}$ one; $\Gamma _{cd}\subset \Gamma
_{0}$. We treat separately the two corresponding  orbifolds
${\cal{O}}_{0}$\ and ${\cal{O}} _{cd}$   and   study their link to
each other.
\\$\mathit{(iii)}$ Finally we construct the
 NC extension of the Calabi-Yau hypersurfaces by
first deriving the right constraint equations,  then solve non
commutativity in terms of discrete torsion and toric  geometry
data of the variety.\\
 This method  can be applied to   higher
dimensional NC  torii orbifolds  extending the result of  NC
$(T^{2}\times T^{2}\times T^{2})/(\mathbf{{Z_{2}}\times {Z_{2}}})$
Calabi-Yau  threefolds.  In this case, the general  solution  is
given in terms of $d$-dimensional Clifford algebra.
\par The organization of this paper is as follows: In section 2,
we review the main lines of Calabi-Yau hypersurfaces using toric
geometry methods. Then  we develop a method of getting complex $d$
Calabi-Yau mirror coset manifolds $C^{k+1}/C^{\ast r}$ , $k-r=d,$
as hypersurfaces
 in $%
WP^{d+1}$, by  solving  the $y_{i}$ inavriants  of mirror geometry
in terms of invariants of the $C^{\ast }$ action of the weighted
projective space and the toric  geometry data of $C^{k+1}/C^{\ast
r}$. In section 3, we explore the general form of discrete
symmetries of the mirror  hypersurface using their toric geometry
data. Then we discuss  orbifolds of toric Calabi-Yau
hypersurfaces. In section 4, we build the corresponding NC toric
Calabi-Yau algebras using the algebraic geometry approach of
\cite{38,39}. Then we work out explicitly the matrix realizations
of these algebras using toric geometry ideas. In section 5, we
give the link with the  BL construction while in section 6 we give
the generalization of the NC $\frac{T^{2}\times
T^{2}\times T^{2}}{\mathbf{{Z_{2}}\times {Z_{2}}}}$ orbifold to $\frac{%
\left( T^{2}\right) ^{\otimes (2k+1)}}{\mathbf{{Z_{2}^{2k}}}}$,
$k\geq 1,$\ where $\left( T^{2}\right) ^{\otimes (2k+1)}$\ is
realized by (2k+1) elliptic curves embedded in $C^{(4k+2)}$
complex space. Our construction, which generalizes that of
\cite{39} given by $k=1$, involves non commuting operators
satisfying the $2k$ dimensional Clifford algebra. We end this
paper by giving our conclusion.
\section{ Toric geometry of CY  Manifolds }

\subsection{Toric  realization of CY manifolds}
The simplest $(d+1)$-complex dimension toric manifold, which we
denote as
$M_{\triangle }^{d+1},$ is given by the usual complex projective space $%
P^{d+1}=\{C^{d+2}-{\bf{0}}_{d+2}\}/C^{\ast }$ \cite{55,56,57}. One
can also
build  $M_{\triangle }^{d+1}$ varieties by considering the $(k+1)$%
-dimensional complex spaces $C^{k+1},$ parameterized by the
complex coordinates
$\{{\bf{x}}=(x_{1},x_{2},x_{3},...,x_{k+1})\},$ and $r$ toric
actions $T_{a}$  acting on the $x_{i}$'s as;
\begin{equation}
T_{a}:x_{i}\rightarrow x_{i}\left( \lambda
_{a}^{q_{i}^{a}}\right).
\end{equation}
Here  the $\lambda _{a}$'s are $r$ non zero complex parameters and
$q_{i}^{a}$  are integers defining the  weights of the toric
actions  $T_{a}$. Under these actions, the $x_{i}$'s form a set of
homogeneous coordinates defining a ($d+1)$ complex dimensional
coset manifold  $M^{d+1}=(C^{k+1})/C^{\ast r}$ with dimension $
d=(k-r)$.\par
More generally, toric manifolds may be thought of as the coset space $%
(C^{k+1}-{\cal P})/C^{\ast r}$with ${\cal P}$ a given subset
 of $C^{k+1}$ defined by the $%
C^{\ast r}$\ action and a chosen triangulation. ${\cal P}$
generalizes  the standard $\{{\bf{0}}_{k+1}=(0,0,0,...,0)\}$
singlet subset that is removed in the case of $P^{k}$. One of the
beautiful features of toric manifolds is their nice geometric
realization known as the toric geometry representation. The toric
data of this realization are\ encoded in a polyhedron $\Delta $
generated by $(k+1)\ $vertices carrying all geometric informations
on the manifold. These data are stable under $C^{\ast r}$ actions
and are useful in the geometric engineering method of $4D\quad $
${\cal N}=2$ supersymmetric quantum field theory in particular in
 the building of the basic $(d+1)$ gauge invariant
coordinates system $\{u_I\}$ of the $(C^{k+1}-{\cal P})/C^{\ast
r}$\ coset manifold in terms of the homogeneous coordinates
$x_{i}$  \cite{50,51,52,53,55}. \par In toric geometry, $(d+1)$
complex manifolds $M_{\triangle }^{d+1}$ are generally represented
by an integral polytope $\Delta $ spanned by $(k+1)$ vertices
${V}_{i}$\ of the standard lattice $Z^{d+1}$. These vertices
fulfill $\ r$ relations given by:
\begin{equation}
\sum_{i=1}^{k+1}q_{i}^{a}{V}_{i}=0,\qquad a=1,...,r,
\end{equation}
and are in one to correspondence with the $r$ actions of $C^{\ast
r}$ on the complex coordinates $x_{i}$ eq(2.1). In the above
relation, the $q_{i}^{a}$ integers are the same as in eq(2.1) and
are interpreted, in the ${\cal N}=2$ gauged  linear sigma model
language, as the $U(1)^{r}$ gauge charges of the $x_{i}$ complex
field variables of two dimensional ${\cal N}=2$ chiral multiplets
[56-62]. They are also known as the entries of the Mori vectors
describing the intersections of complex curves $C_{a}$ and
divisors $D_{i}$ of $M_{\triangle }^{d+1}$ \cite{63,64,65}.
\par Submanifolds $\mathcal{N}$ of \ $M_{\triangle }^{d+1}$ may be
also studied by using the\ $\Delta $ toric data $\{q^a_i,V_i\}$ of
the original manifold. An interesting example of $M_{\triangle
}^{d+1}$\ subvarieties is given by the $d$ complex  dimension
Calabi-Yau manifolds $H_{\triangle }^{d}$ defined as hypersurfaces
in $M_{\triangle }^{d+1}$ as follows \cite{53}:
\begin{equation}
p(x_{1},x_{2},x_{3},...,x_{k+1})=\sum_{I}b_{I}
\prod_{i=1}^{k+1}x_{i}^{<{V}_{i},{V}_{I}^{\ast }>}=0,
\end{equation}
together with the Calabi-Yau condition
\begin{equation}
\sum_{i=1}^{k+1}q_{i}^{a}=0,\qquad a=1,...,r.
\end{equation}
The ${V}_{I}^{\ast }$'s appearing in the relation (2.3) are
vertices in the dual polytope $\Delta ^{\ast }$ of $\Delta ;$
their scalar product with the ${V}_{i}$'s is positive,
$<{V}_{i},{V} _{I}^{\ast }>$ $\geq 0$. For convenience, we will
set  from now on  $<{V} _{i},{V}_{I}^{\ast }>=n_{i}^{I}$.  The
$b_{I}$ coefficients are complex moduli  describing the complex
structure of $H_{\triangle }^{d}$; their number is given by the
 Hodge number $h^{(d-1,1)}(H_{\triangle }^{d})$.
 Using the $n_{i}^{I}$ integers, the $d$ dimensional hypersurfaces $%
H_{\triangle }^{d}$ in $M_{\triangle }^{d+1}$ eq(2.3) read as
\begin{equation}
\sum_{I}b_{I}\prod_{i=1}^{k+1}x_{i}^{n_{i}^{I}}=0.
\end{equation}
At  this stage it is interesting to make some remarks regarding
the above relation . At first sight, one is tempted to make a
correspondence between this relation and the hypersurface eq used
in \cite{39} and take it as the starting point to build  NC
Calabi-Yau manifolds a la  Berenstein et \textit{al}. However this
is not so obvious; first because the polynomial (2.5) is not a
homogeneous one and second even though one wants to try to bring
it to a homogeneous form, one has to specify the toric data
$\{q_{I}^{\ast A};{V}_{I}^{\ast }\}$ of the polyhedron $\Delta
^{\ast }$; mirror to $\{q_{i}^{a};{V}_{i}\}$ data of $\Delta $.
The mirror data satisfy similar relations as (2.2) and (2.4)
namely:
\begin{eqnarray}
\sum_{I=1}^{k^{\ast }+1}q_{I}^{\ast A} &=&0\qquad A=1,...,r^{\ast
},
\\
\sum_{I=1}^{k^{\ast }+1}q_{I}^{\ast A}{V}_{I}^{\ast } &=&0\qquad
A=1,...,r^{\ast }\nn
\end{eqnarray}
 together with $ k+1-r =k^{\ast }+1-r^{\ast }=d$.
  Moreover setting $Y_{I}=\Pi
_{i=1}^{k+1}$\ $x_{i}^{n_{i}^{I}},$ the above polynomial becomes a
linear combination of the $Y_{I}$\ gauge invariants as
$\sum_{I}b_{I}\ Y_{I}=0.$ This relation can however be rewritten
in terms of the $(d+1)$ dimensional generator basis $\{Y_{\alpha
};1\leq \alpha \leq (d+1)\}$ as follows
\begin{equation}
1+\sum_{\alpha =1}^{d+1}b_{\alpha } Y_{\alpha
}+\sum_{I=d+2}^{k^{\ast }+1}b_{I}\ Y_{I}=0,
\end{equation}
where the remaining $Y_{I}$ invariants$,$ that is the set
$\{Y_{I};(d+2)\leq I\leq (k^{\ast }+1)\}$\ are determined by
solving the following Calabi-Yau constraint eqs
\begin{equation}
\prod_{I=1}^{k^{\ast }+1}Y_{I}^{q_{I}^{\ast A}}=1;\quad
A=1,...,r^{\ast }.
\end{equation}
To realize the relation (2.7) as a homogeneous polynomial
describing the hypersurfaces $H_{\triangle }^{d}$ with the desired
properties, in particular the Calab-Yau condition, one has to
solve the above constraint eqs. Though this derivation can apriori
be done using (2.8), we will not proceed  in that way. What we
will do instead is to use the so called mirror Calabi-Yau
manifolds $H_{\triangle }^{d\ast }$ and derive its homogeneous
description. The point is that the mirror geometry has some
specific features and constraint eqs that involve directly  the
toric data $\{q_{i}^{a};{V}_{i}\}$\ of the $\Delta $ polyhedron
contrary to the original hypersurfaces $H_{\triangle }^{d}$ which involve the $%
\{q_{I}^{\ast A};{V}_{I}^{\ast }\}$\ data of $\Delta ^{\ast }$.
Once the rules of getting the $H_{\triangle }^{d\ast }$
homogeneous hypersurfaces are defined, one can also reconsider the
analysis of $H_{\triangle }^{d}$ by starting from the relations
(2.7-8), use the $\Delta ^{\ast }$ toric data and perform similar
analysis to that we will be developing herebelow.
\par Under mirror symmetry, toric manifolds $M_{\triangle
}^{(d+1)}$ and Calabi-Yau hypersurfaces $H_{\triangle }^{d}$ are
mapped to $M_{\triangle }^{(d+1)\ast }$ and $H_{\triangle }^{d\ast
}$ respectively. They are obtained by exchanging the roles of
complex and Kahler structures in agreement with the Hodge
relations
\begin{eqnarray}
h^{(d-1,1)}(H_{\triangle }^{d}) &=&h^{(1,1)}(H_{\triangle }^{d\ast }),
 \\
h^{(1,1)}(H_{\triangle }^{d}) &=&h^{(d-1,1)}(H_{\triangle }^{d\ast
})\nn,
\end{eqnarray}
and similarly for $M_{\triangle }^{(d+1)}\ $ and
 $M_{\triangle }^{(d+1)\ast }$ [64-67]. In practice the building of $M_{\triangle }^{(d+1)\ast }$ and so $%
H_{\triangle }^{d\ast }$ is achieved by using the vertices ${V}%
_{I}^{\ast }$\ of the\ convex hull spanned by the ${V}_{\alpha
}^{\ast }$. Following \cite{66,67,68,69,70,71,72},  mirror
Calabi-Yau manifolds $H_{\triangle }^{d\ast }$ is given by the
zero of the polynomial
\begin{equation}
p(z_{1},z_{2},...,z_{k^{\ast }+1})=\sum_{i=1}^{k+1}a_{i}%
\prod_{I=1}^{k^{\ast }+1}\left( z_{I}^{n_{i}^{I}}\right) ,
\end{equation}
where the $z_{I}$'s are the mirror coordinates. The $C^{\ast
r^{\ast }}$ actions of $M_{\triangle }^{(d+1)\ast }$ act on the
$z_{I}$'s as
\begin{equation}
z_{I}\rightarrow z_{I}\lambda _{I}^{q_{I}^{\ast A}},
\end{equation}
with $q_{I}^{\ast A}$\ as in eq(2.6). The $a_{i}$'s are the
complex structure of the mirror Calabi-Yau manifold $H_{\triangle
}^{d\ast };$ they describe also the Kahler deformations of
$H_{\triangle }^{d}$.  An interesting feature of the relation
(2.10) is its representation in terms of the $(k+1)$ invariants
$y_{i}=\Pi _{I=1}^{k^{\ast }+1}$\ $\left(
z_{I}^{m_{i}^{I}}\right) $ under the $C^{\ast r^{\ast }}$ actions of $%
M_{\triangle }^{d\ast }$; i.e:
\begin{equation}
\sum_{i=1}^{k+1}a_{i}y_{i}=0,
\end{equation}
together with the $r$ following constraint eqs of the mirror
geometry
\begin{equation}
\prod_{i=1}^{k+1}\left( y_{i}^{q_{i}^{a}}\right) =1,\qquad
a=1,...,r.
\end{equation}
 These eqs involve $(k+1)$ variables $y_{i}$, not all of them
independent since they are subject to $(r+1)$ conditions ( $r$
from eqs(2.13) and one from (2.12)) leading indeed to the right
dimension of $H_{\triangle }^{d\ast }$. Eqs(2.12-13) will be our
starting point towards building  NC alabi-Yau manifolds using the
Berenstein et \textit{al} approach. Before that let us put these
relations into a more convenient form.

\subsection{Solving the mirror  constraint eqs }
As shown on the above eqs, not all the $y_{i}$'s are independent
variables, only $(d+1)$ of them do. In what follows we shall fix
this redundancy by using\ a coordinate patch of the (d+1) weighted
projective spaces $WP^{d+1}$ parameterized by the system of
variables $\left\{ u_{\alpha },1\leq \ \alpha
\leq (d+1);u_{d+2}\right\}$. In the coordinate patch $u_{d+2}=1,$\ the $%
u_{\alpha }$ variables behave as $(d+1)$\ independent gauge invariants
parameterizing the coset manifold $\left[ (C^{d+2})/C^{\ast }\right] \sim %
\left[ (C^{k+1})/C^{\ast r}\right] $. The $r$ remaining $y_{i}$'s
\ are given by monomials of the $u_{\alpha }$'s. A nice way of
getting the relation  between   $y_{i}$'s and  $u_{\alpha }$'s is
inspired from the analysis \cite{53, 54}; it is based on
introducing the following system $\left\{ N_{i};1\leq
\ i\leq (k+1)\right\} $\ of $(d+1)$ dimensional vectors of integer entries $%
\left( N_{i}\right) _{\alpha }=<{V}_{i},{V}_{\alpha }^{\ast
}>\equiv n_{i}^{\alpha }$. From eq(2.2), it is not difficult to
see that:
\begin{equation}
\sum_{i=1}^{k+1}q_{i}^{a}N_{i}=0,\qquad a=1,...,r;\qquad \alpha
=1,...,d+1,
\end{equation}
or equivalently:
\begin{equation}
\sum_{i=1}^{k+1}q_{i}^{a}n_{i}^{\alpha }=0,\qquad a=1,...,r;\qquad
\alpha =1,...,d+1.
\end{equation}
Note that the introduction of the system $\left\{ \left(
N_{i}\right) _{\alpha }\equiv n_{i}^{\alpha };1\leq \ i\leq
(k+1)\right\} $\ has a
remarkable interpretation; it describes the complex deformations of $%
H_{\triangle }^{d\ast }$\ and by the correspondence (2.9) the Kahler ones of $%
H_{\triangle }^{d}$. Observe also that shifting the $N_{i}$'s by a
constant vector, say $t_{0}$, eq(2.14) remains invariant due to
the Calabi-Yau condition (2.4). Therefore the $V_{i}$\ vertices of
eqs(2.2) can
be solved by a linear combination of $N_{i}$\ and $t_{0};$ $%
V_{i}=N_{i}+at_{0}.$ Having these relations in mind, we\ can use them to
reparametrize the $y_{i}$ invariants in terms of the $(d+2)$ generators $%
u_{\mu }$\ $(u_{d+2}$ arbitrary) as follows
\begin{eqnarray}
y_{i}
&=&u_{1}^{(n_{i}^{1}-1)}u_{2}^{(n_{i}^{2}-1)}...u_{d+1}^{(n_{i}^{d+1}-1)}
u_{d+2}^{(n_{i}^{d+2}-1)}=\prod_{\mu =1}^{d+2}u_{\alpha }^{(n_{i}^{\mu }-1)},
\\
y_{0} &=&1\qquad \Leftrightarrow \qquad (n_{0}^{\alpha }-1)=0,\qquad \forall
\alpha =1,...,d+2.
\end{eqnarray}
Note that $\Pi _{i=1}^{k+1}(y_{i}^{q_{i}^{a}})=1$ is automatically
satisfied due to eqs(2.14) and(2.15). Note also the $n_{i}^{d+2}$
integers are extra quantities introduced for later use; they
should not be confused with the $\left\{ n_{i}^{\alpha };1\leq
\alpha \leq d+1\right\} $ entries of $N_{i}$. Putting the
relations (2.16-17) back into eq(2.12), we get an equivalent way
of writing eq(2.10), namely:
\begin{equation}
a_{0}1+\sum_{i=1}^{k+1}a_{i}%
u_{1}^{(n_{i}^{1}-1)}u_{2}^{(n_{i}^{2}-1)}...u_{d+1}^{(n_{i}^{d+1}-1)}
u_{d+2}^{(n_{i}^{d+2}-1)}=0.
\end{equation}
The main difference between this relation and eq(2.10) is that the
above one involve $(d+2)$ variables only,  contrary to the case of
eq (2.10) which rather involve $(d+r^{\ast }+1)$ coordinates; that
is $r^{\ast }$ variables in more. Eq (2.18) is then  a relation
where the $C^{\ast r^{\ast }}$ symmetries on the $z_{I}$'s
eq(2.11) are completely fixed. Indeed starting from eq (2.10), it
is not difficult to rederive eq (2.18) by working in the
remarkable coordinate patch ${\cal U}=\left\{
(z_{1},z_{2,}...,z_{d+2,}1,1,...,1)\right\} ,$ which is isomorphic
to a weighted projective space $WP_{(\delta _{1},...,\delta
_{d+2})}^{d+1}$ with a weight vector ${\delta }_{\mu }=(\delta
_{1},...,\delta _{d+2})$. In this way of viewing  things, the
$y_{i}$\ variables may be thought of as gauge invariants under the
projective  action $WP_{(\delta _{1},...,\delta _{d+2})}^{d+1}$\
 and consequently the Calabi-Yau manifold (2.18) as a hypersurface in $%
WP_{(\delta _{1},...,\delta _{d+2})}^{d+1}$ described by a
homogeneous polynomial $p(u_{1},...,u_{d+2})$ embedded  of degree
$D=\sum_{\mu =1}^{d+2}\delta _{\mu }$.\ Thus, under the projective
action $u_{\mu }\longrightarrow \lambda ^{\delta _{\mu }}$ $u_{\mu
},$ the monomials\ $y_{i}=\Pi _{\mu
=1}^{d+2}\left( u_{\mu }^{(n_{i}^{\mu }-1)}\right) $ transform as \ $%
y_{i}\lambda ^{\Sigma _{\mu }\left( \delta _{\mu }(n_{i}^{\mu
}-1)\right) }$ and so    the following  constraint eqs should
hold,
\begin{eqnarray}
\sum_{\mu =1}^{d+2}\delta _{\mu } &=&D, \\
\sum_{\mu =1}^{d+2}\delta _{\mu }n_{i}^{\mu } &=&D.
\end{eqnarray}
These relations show that the $n_{i}^{\mu }$\ integers can be
solved in terms of the partitions$\ d_{i}^{\mu }$\ of the degree
$D$ of the homogeneous
polynomial $p(u_{1},...,u_{d+2})$. Indeed from  $%
\sum_{\mu =1}^{d+2}d_{i}^{\mu }=D,$ one sees that $n_{i}^{\mu }=\frac{%
d_{i}^{\mu }}{\delta _{\mu }};$ among which we have the following
remarkable ones
\begin{equation}
n_{i}^{\mu }=\frac{D}{\delta _{\mu }}\qquad if\qquad i=\mu \quad
\mbox{for} 1\leq \mu \leq d+2.
\end{equation}
To get the $V_{i}$ vertices, we keep the $\left\{ n_{i}^{\alpha
};1\leq \alpha \leq d+1\right\} $ entries and substruct the
trivial monomial associated with  $\left\{ (t_{0}^{\alpha
})=(1,1,...,1)\right\} $. So  the  $V_i$ vertices are
\begin{equation}
V_{i}^{\alpha }=n_{i}^{\alpha }-t_{0}^{\alpha }=\frac{d_{i}^{\alpha }}{%
\delta _{\alpha }}-t_{0}^{\alpha }.
\end{equation}
For the$\ (d+3)$ leading vertices, we have:
\begin{eqnarray}
V_{0} &=&(0,0,0,...,0,0)  \nn\\ V_{1} &=&(\frac{D}{\delta
_{1}}-1,-1,-1,...,-1,-1),  \nn \\ V_{2} &=&(-1,\frac{D}{\delta
_{2}}-1,-1,...,-1,-1)  \\ V_{3} &=&(-1,-1,\frac{D}{\delta
_{3}}-1,...,-1,-1) \nn\\ &&...   \nn\\ V_{d+1}
&=&(-1,-1,-1,...,\frac{D}{\delta _{d+1}}-1,-1),   \nn\\ V_{d+2}
&=&(-1,-1,-1,...,-1,\frac{D}{\delta _{d+2}}-1).\nn
\end{eqnarray}
Before going ahead, let us give some remarks: $(a)$ the
integrality of the entries of these vertices requires that the $D$
degree should be a commun multiple of the  weights $\delta _{\mu
}$. Moreover the number  of partitions of  $D$ should be less than
$(k+2).$ $(b)$ As far the ($d+3$) leading vertices are concerned,
the corresponding homogeneous monomials are
\begin{eqnarray}
N_{0} &\rightarrow &\prod_{\mu =1}^{d+2}u_{\mu }, \\ N_{\mu }
&\longrightarrow &u_{\mu }^{\frac{D}{\delta _{\mu }}},\quad  \mu
=1,...,d+2.
\end{eqnarray}
So the corresponding mirror polynomial takes the form:
\begin{equation}
\sum_{\mu =1}^{d+2}u_{\mu }^{\frac{D}{\delta _{\mu }}}+a_{0}\prod_{\mu
=1}^{d+2}\left( u_{\mu }\right) =0.
\end{equation}
More generally the mirror polynomial $P_{\Delta }(u)$\ describing $%
H_{\triangle }^{d\ast }$\ reads as\
\begin{equation}
P_{\Delta }(u)=\sum_{\mu =1}^{d+2}u_{\mu }^{\frac{D}{\delta _{\mu }}%
}+a_{0}\prod_{\mu =1}^{d+2}\left( u_{\mu }\right) +\sum_{i=d+3}^{k+1}a_{i}%
\prod_{\mu =1}^{d+2}(u_{\mu }^{n_{\mu }^{i}})=0,
\end{equation}
where the $a_{i}$'s  are complex moduli of the mirror Calabi-Yau
hypersurface.
\subsection{More on the  mirror CY geometry }
Here we further explore the relations between the realizations
(2.10) and (2.27) of the mirror Calabi-Yau manifolds. In
particular, we give an explicit derivation of the weights $\delta
_{\mu }$\ involved in the polynomials (2.27) in terms of the
Calabi-Yau $q_{i}^{a}$ charges. To do so, first of all recall that
under the projective action
\begin{equation}
u_{\mu }\longrightarrow \lambda ^{\delta _{\mu }}u_{\mu },
\end{equation}
the polynomial $P_{\Delta }(u)$ behaves as $P_{\Delta }(\lambda
^{\delta _{\mu }}u)=\lambda ^{D}P_{\Delta }(u)$ leaving the zero
locus invariant. Using the identity $\sum_{\mu =1}^{d+2}\delta
_{\mu }=D$, one may reinterpret the Calabi-Yau condition (2.4) or
equivaletly by introducing $r$
integers $p_{a} $%
$$ \sum_{\mu =1}^{d+2}\sum_{a=1}^{r}p_{a}q_{\mu
}^{a}=-\sum_{i=d+3}^{k+1}\sum_{a=1}^{r}p_{a}q_{i}^{a}, $$ by
thinking about it as
\begin{eqnarray}
\delta _{\mu } &=&\sum_{a=1}^{r}p_{a}q_{\mu }^{a} \\ D
&=&\sum_{i=d+3}^{k+1}\delta
_{i}=-\sum_{i=d+3}^{k+1}\sum_{a=1}^{r}p_{a}q_{i}^{a}.
\end{eqnarray}
 For instance, for ordinary projective spaces $ P^k$,  we  can use the
generalization of the transformation introduced in \cite{39};
namely \bea u_{\mu }&\longrightarrow & \omega ^{Q_{\mu
}^{a}}u_{\mu },\nn\\u_{d+2}&\longrightarrow & u_{d+2}, \eea where
,roughly speaking,  $\omega $ is a $D-th$ root of unity. This
transformation leaves $P_{\Delta }(u)$\ invariant as far as the
$Q_{\mu }^{a} $'s obey the Calabi-Yau condition $\Sigma _{\mu
=1}^{d+1}Q_{\mu }^{a}=0$ and $Q_{d+2}^{a}=0$, in agreement with
the choice of the coordinate patch $u_{d+2}=1$. Next by
appropriate choice of $\lambda $, we can compare both the
transformations(2.28) and (2.31) as well as their actions on the
 monomials $%
y_{i}=\Pi _{\mu =1}^{d+2}\left( u_{\mu }^{(n_{i}^{\mu }-1)}\right)
$ respectively given by $y_{i}\longrightarrow  y_{i}\omega
^{\Sigma _{\mu }\delta _{\mu }(n_{i}^{\mu }-1)}\ $ and
$y_{i}\longrightarrow y_{i}$ $\omega ^{\Sigma _{\mu }Q_{\mu
}^{a}(n_{i}^{\mu }-1)}. $\ Invariance under these actions lead  to
eqs (2.19-20), and their toric  geometry equations analogue
\begin{eqnarray}
\sum_{\mu =1}^{d+2}Q_{\mu }^{a} &=&0\qquad  \mbox {modulo (D) }
\\
\sum_{\mu =1}^{d+2}Q_{\mu }^{a}n_{i}^{\mu } &=&0;\quad \mbox
{modulo (D) }.
\end{eqnarray}
 Comparing these eqs  with eqs(2.32-33) and (2.19-20), one
gets the following relation between the $Q_{\mu }^{a}$ and
$q_{i}^{a}$ charges of the original manifold
\begin{equation}
Q_{\mu }^{a}=\left( q_{\mu }^{a}+\frac{{1}}{d+2}%
\sum_{i=d+3}^{k}q_{i}^{a}\right) ;\quad \mbox {modulo (D) }
\end{equation}
As the isometries of eqs(2.26-27) will be involved in the study of
the NC hypersurface Calabi-Yau orbifolds, let us derive  general
form of these isometries using geometry toric data. We will
distinguish  between two cases: (i) the group of isometries
$\Gamma _{0}$ leaving eq(2.26) invariant. (ii)   its subgroup
$\Gamma _{cd}$ of discrete symmetries of eq(2.27). commuting  with
complex deformations.
\section{Discrete Symmetries and   CY Orbifolds }
To determine the discrete symmetries of the  Calabi-Yau
homogeneous hypersurfaces, let us derive  the general groups
$\Gamma _{0}$\ and $\Gamma _{cd} $ of transformations leaving
eqs(2.26) and (2-27) invariants:
\begin{equation}
\Gamma =\{g_{\omega }\mid g_{ W }:u_{\mu }\rightarrow g_{\omega
}(u_{\mu })=u_{\mu }^{\prime }=u_{\mu }\left( { \cal W}%
\right) ^{{\bf b}_{\mu }};\quad P_{\Delta }(u^{\prime })=P_{\Delta
}(u)\quad \},
\end{equation}
where $${\cal W}^{{\bf b}_{\mu }}=\Pi _{\nu =1}^{d+2}\left[ \left(
\omega _{\nu }\right)^{{\bf a}_{\mu }^{\nu }}\right] $$ and where
$\left\{ {\bf b}_{\mu }\right\} _{1\leq \mu \leq d+2}$ is a
$(d+2)$ dimensional vector weight and $\bf a_{\mu }^{\nu }$\ are
their entries. They will be determined by symmetry requirements
and the Calabi-Yau toric geometry  data. As the solutions we will
build depend on the weights $\delta _{\mu }$, we  will distinguish
hereafter the $P^{d+1}$ and $WP^{d+1}$ spaces; a matter of
illustrating the idea and the techniques we will be using.

\subsection{$P^{d+1}$ projective spaces}

The crucial point to note here is that because of the equality
$\delta
_{1}=\delta _{2}=...=\delta _{d+2}=1$, the $D$\ degree of the polynomials $%
P_{\Delta }(u)$ is equal to $(d+2)$ and so the constraint eq
(2.20) reduces
to: $\sum_{\mu =1}^{d+2}n_{i}^{\mu }=(d+2)$ for any value of the $%
i $ index. Putting back $\delta _{\mu }=1$ in eqs (2.26), one sees
that invariance under $\Gamma _{0}$ of the first terms $u_{\mu
}^{d+2}$ shows that a natural solution is given by taking $\omega
_{1}=\omega _{2}=...=\omega _{d+1}=\omega =\exp i(\frac{2\pi
}{d+2})$ and then $\omega^{{\bf b}_{\mu }}=$ $\exp i\frac{2\pi
}{d+2}{\bf b}_{\mu }$.  However, invariance of the term
$\prod_{\mu =1}^{d+2}($ $u_{\mu })$ under the change (3.1),
implies that $ {\bf b}_{\mu }$\ should satisfy the following
constraint equation
\begin{equation}
\sum_{\mu =1}^{d+2}{\bf b}_{\mu }=0,\qquad \mbox {modulo (d+2) }.
\end{equation}
In what follows, we shall give an explicit class of special
solutions for the constraint eq $\sum_{\mu =1}^{d+2}{\bf b}_{\mu
}=0$, by using the toric geometry  data of the $H_{\triangle
}^{d}$\ Calabi-Yau manifold eqs(2.2) and (2.4). The solutions,
modulo $(d+2)$, are obtained by making appropriate shifts.

\subsubsection{Explicit construction of ${\bf b}_{\protect\mu }$\ weights\ }
The solution for $b_{\mu }$ we will construct herebelow contains
two terms which are intimately linked to  toric geometry eqs (2.2)
and (2.4). To have an idea on the explicit derivation of the ${\bf
b} _{\mu }$'s, let us first introduce the two following $Q_{\mu }$
and $\xi _{\mu }$ quantities. They will be used in \ realizing
${\bf b}_{\mu }$.\\ {\bf The $Q_{\protect\mu }$Weights :}\\ This
is a quantity defined as:
\begin{equation}
Q_{\mu }=Q_{\mu }(p_{1},...,p_{r})=\sum_{a=1}^{r}p_{a}Q_{\mu }^{a},\qquad
1\leq \mu \leq d+2,
\end{equation}
 where the $p_{a}\ ^{\prime }s$ are given integers and where
 $Q_{\mu }^{a}$ are a kind of shifted Calabi-Yau charges; they are  given in
terms of the $q_{\mu }^{a}$\ Mori vectors of the toric manifold
shifted by constant numbers $\tau ^{a};$ as shown on the following
relation
\begin{equation}
Q_{\mu }^{a}=q_{\mu }^{a}+\tau ^{a}.
\end{equation}
The $\tau ^{a}$'s are determined by requiring that the $Q_{\mu
}^{a}$\ shifted charged have  to satisfy  the Calabi-Yau condition
$\sum_{\mu =1}^{d+2}Q_{\mu }^{a}=0$. Using (2.4), we find
\begin{equation}
\tau ^{a}=\frac{{1}}{d+2}\sum_{i=d+3}^{k+1}q_{i}^{a}.
\end{equation}
Replacing $Q_{\mu }^{a}$\ by its explicit expression in terms of
the Mori vector charges, we get
\begin{equation}
Q_{\mu }=\sum_{a=1}^{r}p_{a}\left( q_{\mu }^{a}+\frac{{1}}{d+2}%
\sum_{i=d+3}^{k}q_{i}^{a}\right).
\end{equation}
It satisfies identically the property $\sum\limits_{\mu
=1}^{d+2}Q_{\mu }=0,$ which we will interpret  as the Calabi-Yau
condition because of its link with the original relation
$\sum\limits_{i=1}^{k+1}q_{i}^{a}=0.$\\ {\bf The $\protect\xi
_{\protect\mu }$ Weights :}\\ These weights carry informations on
the data of the polytope $\Delta $ of the toric varieties and so
on their Calabi-Yau submanifolds. They are defined as
\begin{equation}
\xi _{\mu }=\xi _{\mu }(s_{1},...,s_{d+1})=\sum_{\alpha =1}^{d+1}s_{\alpha
}\xi _{\mu }^{\alpha }
\end{equation}
where the $s_{\alpha }$'s are integers and where $\xi _{\mu
}^{\alpha }$ are defined in terms of the toric data of
$M_{\triangle }^{d+1}$\ as follows
\begin{equation}
\xi _{\mu }^{\alpha }=\sum_{a=1}^{r}p_{a}\left( q_{\mu }^{a}n_{\mu
}^{\alpha }+\frac{1}{d+2}\sum_{i=d+3}^{k+1}q_{i}^{a}n_{i}^{\alpha
}\right).
\end{equation}
Like for the $Q_{\mu }$ weights, one can check  here also that the
sum  $\sum\limits_{\mu =1}^{d+2}\xi _{\mu }$ vanishes identically
due to the constraint equation(2.15).\\ {\bf The $ {\bf
b}_{\protect\mu }$\ Weights :}\\ A class of solutions for $\bf
b_{\mu }$ based on the Calabi-Yau toric geometry  data (2.2) and
(2.4), may be given by a \ linear combination of the $Q_{\mu }$
and $\xi _{\mu }$ weights as shown herebelow:
\begin{equation}
{\bf b}_{\mu }=m_{1}Q_{\mu }+m_{2}\xi _{\mu },
\end{equation}
where $m_{1}$and $m_{2}$ are integers modulo $(d+2)$.   Moreover
setting $ {\bf b}_{\mu }=\sum_{\nu =1}^{d+2}$ ${\bf a}_{\mu }^{\nu
}$ and
\begin{eqnarray}
Q_{\mu }^{\alpha } &=&Q_{\mu }^{a}\qquad for\qquad \alpha =1,...,r
; \nn \\ Q_{\mu }^{\alpha } &=&0\qquad for\qquad \alpha
=(r+1),...,(d+2),
\end{eqnarray}
while $Q_{\mu }^{\alpha }=Q_{\mu }^{a}$\ for $r\geq d+1,$ we can
rewrite the above solutions as follows:
\begin{equation}
{\bf a}_{\mu }^{\nu }=m_{1}Q_{\mu }^{\nu }+m_{2}\xi _{\mu }^{\nu
}.
\end{equation}
Therefore,  the general transformations of the $\Gamma
_{0}(P^{d+1})$ group of discrete isometries are given by the
change (3.1) with ${\bf b}_{\mu }$ vector
weights depending on $(r+d+1)=k$ integers; namely $r$ integers $p_{a}$\ and $%
(d+1)$\ integers $s_{\alpha }.$
\subsubsection{Complex Deformations }
To get the discrete symmetries of the full Calabi-Yau homogeneous
complex hypersurface including the complex deformations eq(2.27),
one should solve more complicated constraint relations which we
give hereafter. Under $\Gamma _{cd}$ of transformations eq(2.27),
the complex deformations of the Calabi-Yau manifold $P_{\Delta
}(u)$\ are stable provided the ${\bf b}_{\mu }$\ weights satisfy
eq(3.2) but also the following constraint eqs:
\begin{equation}
\sum_{\mu =1}^{d+2} {\bf b}_{\mu }n_{\mu }^{\nu }=0,
\end{equation}
where the $n_{\mu }^{\nu }$ 's\ are as in eq(2.27).\ A particular
solution
of these constraint eqs is given by taking ${\bf b}_{\mu }=Q_{\mu }$\ that is $%
m_{1}=1$ and $m_{2}=0$. Indeed replacing ${\bf b}_{\mu }$\ by its
expression(3.9) and putting back into the above relation, we get
by help of the identity (2.20),
\begin{eqnarray}
\left[ \sum_{\mu =1}^{d+2}\sum_{a=1}^{r}p_{a}(q_{\mu }^{a}+\tau
^{a})n_{\mu }^{\nu }\right] &=&\sum_{a=1}^{r}p_{a}\left[ \sum_{\mu
=1}^{d+2}q_{\mu }^{a}n_{\mu }^{\nu }+(d+2)\tau ^{a}\right] \nn\\
&=&\sum_{a=1}^{r}p_{a}\left[ \sum_{\mu =1}^{d+2}q_{\mu }^{a}n_{\mu
}^{\nu }+\sum_{i=d+3}^{k}q_{i}^{a}n_{\mu }^{\nu }\right] =0.
\end{eqnarray}
For $m_{1}, m_{2}\neq 0,$\ the relation ${\bf b}_{\mu
}=m_{1}Q_{\mu }+m_{2}\xi _{\mu }$ cease to be a solution of the
constraint eq(3.12). Therefore $\Gamma _{cd} $\ is a subgroup of
$\Gamma _{0}$. It depends on the $p_{a}$\ integers and involves
the Calabi-Yau condition only.

\subsection{$WP^{d+1}$ weighted projective spaces}
The previous analysis made for the case of $P^{d+1}$ applies as well for $%
WP^{d+1}.$ Starting from eq(2.26) and making the change (3.1),
invariance requirement leads to take the $\omega _{\mu }$\ group
parameters as $\omega _{\mu }=\exp i\frac{2\pi \delta _{\mu }}{D}$
and the ${\bf a}_{\mu }^{\nu }$\ coefficients constrained as \bea
\sum_{\nu =1}^{d+2}\delta _{\nu }{\bf a}_{\mu }^{\nu }&=&0,\quad
\mbox{modulo} \; \delta_\mu \nn\\
 \sum_{\mu
=1}^{d+2}{\bf a}_{\mu }^{\nu }&=&0. \eea Following the same
reasoning as before, one can write down a class of solution, with
integer entries, in terms of the previous weights as follows
\begin{equation}
{\bf a}_{\mu }^{\nu }=\left( \delta ^{\nu }\right) ^{-1}\left[
m_{1}Q_{\mu }^{\nu }+m_{2}\xi _{\mu }^{\nu }\right],
\end{equation}
where\ $Q_{\mu }^{\nu }$\ and \ $\xi _{\mu }^{\nu }$ are as in eq
(3.11). In case where the complex deformations of eq(2.27) are
taken into account, the discrete symmetry group is no longer the
same since the constraint eq(3.13) is now replaced by the
following one
\begin{equation}
\sum_{\mu =1}^{d+2}{\bf a}_{\mu }^{\nu }n_{\mu }^{i}=0, \forall
\quad \nu =1,...,(d+2).
\end{equation}
Like in the projective case where the $\delta _{\mu }$'s are equal
to one,
the solutions for the ${\bf a}_{\mu }^{\nu }$\ integers are given  by eq(3.15) with $%
m_{1}\neq 0$\ and $m_{2}=0$. To conclude this section, one should
note that the group of discrete isometries $\Gamma _{cd}\subset
\Gamma _{0}$ of the Calabi-Yau hypersurfaces including complex
deformations is intimately related to the Calabi-Yau condition.

\section{NC  Toric CY  Manifolds}
Before exposing our results regarding NC
 toric Calabi-Yau's, let us begin this section by reviewing briefly the
 BL  idea of building NC orbifolds of Calabi-Yau hypersurface.

\subsection{Algebraic geometric  approach for CY  }

Roughly speaking,  given a  $d$ dimensional  a Calabi-Yau manifold
$X^{d}$ described algebraically by  a complex equation $p(z_i)=0$
with a group $\Gamma $ of discrete isometries. Taking the quotient
of $X^{d},$ by  the action of the finite group $\Gamma$
 \be
 \Gamma :\quad z_i\to gz_ig^{-1},\quad g\in \Gamma
 \ee
 such that the two following conditions are fulfilled  $p(z_i)$ polynomial and
  the $(d,0)$ holomorphic    from  are invaraints.  The latter condition  is equivalently
   to the vanishing   the first Chern class  $ c_1=0$. Using the  discrete
   torsion, one can build the NC extensions of the
 orbifold, $\left({{{{X}}^d \o \Gamma}}\right)_{nc}$ as follows.
 The coordinate  $z_{i}$'s are replaced by matrix operators $Z_i$ satisfying
\be
Z_iZ_j=\theta_{ij}Z_jZ_i,
 \ee
 Invariance  of  $p(z_i)$  requires that  the parameter
 $\theta_{ij}$'s to be in the  discrete group $ \Gamma$. Moreover, the
 Calabi-Yau condition   imposes the  extra constraint  equation
 \be
  \prod_i\theta_{ij}=1, \quad \forall j\neq i,
 \ee
In this case of  the quintic, embedded in a $P^{5}$ projective
space described by  the homogeneous polynomial
$p(z_{1},...,z_{5})$ of  degree 5:
\begin{equation}
 p(z_i)=z_{1}^{5}+z_{2}^{5}+z_{3}^{5}+z_{4}^{5}+z_{5}^{5}+a_{0}\prod_{1
=1}^{5}z_{i }=0.
\end{equation}
The group $\Gamma $ acts as $z_{i }\longrightarrow z_{i }$ $%
\omega ^{Q_{i }^{a}}$ where  $\omega^5=1$ and where the $Q_{i
}^{a}$ vectors are
\begin{eqnarray}Q_{i }^{1} &=&(1,-1,0,0,0), \nn \\ Q_{i }^{2}
&=&(1,0,-1,0,0) , \\ Q_{i }^{3} &=&(1,0,0,-1,0)\nn.
\end{eqnarray}
 In the coordinate patch
${\cal U}=\left\{ (z_{1},z_{2},z_{3},z_{4});z_{5}=1\right\} $,
eq(4.5) reduces to
\begin{equation}
1+z_{1}^{5}+z_{2}^{5}+z_{3}^{5}+z_{4}^{5}+a_{0}\prod_{j
=1}^{4}z_{j }=0.
\end{equation}
The local NC  algebra ${\cal {A}}_{nc}$\ describing the NC version
of equation(4.5) is obtained by associating to $z_{5}$\ the matrix
$z_{5}I_{5}$\ and to each holomorphic variable $z_{i}$\ a \
$5\times 5$\ matrix $Z_{i}$ satisfying the BL algebra
\begin{eqnarray}
Z_{1}Z_{2} &=&\alpha Z_{2}Z_{1},\qquad Z_{1}Z_{3}=\alpha
^{-1}\beta Z_{3}Z_{1},   \nn\\ Z_{1}Z_{4} &=&\beta
^{-1}Z_{4}Z_{1},\qquad Z_{2}Z_{3}=\alpha \gamma Z_{3}Z_{2}, \\
Z_{2}Z_{4} &=&\gamma ^{-1}Z_{4}Z_{2},\qquad Z_{3}Z_{4}=\beta
\gamma Z_{4}Z_{3}\nn
\end{eqnarray}
where $\alpha ,\beta $ and $\gamma $ are fifth roots of unity. The
centre  of this algebra  ${\cal{Z(A}}_{nc})=\left\{ I_{5},Z_{\nu
}^{5},\Pi _{\nu =1}^{4}Z_{\nu }\right\} $, that is,
\begin{eqnarray}
\left[ Z_{\mu },Z_{\nu }^{5}\right] &=&0,  \nn\\ \left[ Z_{\mu
},\Pi _{\nu =1}^{4}Z_{\nu }\right] &=&0,
\end{eqnarray}

According to the Schur lemma, one can set $Z_{\nu
}^{5}=I_{5}z_{\nu }^{5}$ and $\Pi _{\nu =1}^{4}Z_{\nu }=I_{5}\Pi
_{\nu =1}^{4}z_{\nu }$ and so the centre coincide with the
equation of the quintic. In what follows we extend  this analysis
to  NC toric Calabi-Yau orbifolds.

\subsection{NC toric CY orbifolds}

Following the same lines as \cite{38,39,40,41,42} and using the
discrete symmetry group $\Gamma
$, one can build the orbifolds  ${\cal{O}}=H_{\triangle }^{d\ast }/$ $%
\Gamma $ of the Calabi Yau hypersurface and work out their non
commutative
extensions ${\cal{O}}_{nc}$. The main steps in the building of ${\cal{O}}%
_{nc}$ may summarized as follows: First start from the Calabi-Yau
hypersurfaces $H_{\triangle }^{d\ast }$ eqs(2.26-27) and fix a
coordinate pach of $WP^{d+1}$, say $u_{d+2}=1$. Then impose the
identification under the discrete automorphisms (3.1) defining
 $H_{\triangle }^{d\ast }/\Gamma $. The NC  extension of this orbifold is obtained as usual by
extending the commutative algebra ${\cal{A}}_{c}$\ of
functions on $H_{\triangle }^{d\ast }/$ $\Gamma $ to a NC one $%
{\cal{A}}_{nc}\sim {\cal{O}}_{nc}.$ In this algebra, the $u_{\mu
}$ coordinates are replaced by matrix  operators $U_{\mu }$\
satisfying the algebraic relations
\begin{equation}
U_{\mu }U_{\nu }=\theta _{\mu \nu }U_{\nu }U_{\mu },\qquad \nu
>\mu =1,...,d+1,
\end{equation}
where the $\theta _{\mu \nu }$ non commutativity parameters obey
the following constraint relations
\begin{eqnarray}
\theta _{\mu \nu }\theta _{\nu \mu } &=&1 , \\ \left( \theta _{\mu
\nu }\right) ^{\frac{D}{\delta _{\nu }}} &=&1 , \\ \prod_{\mu
=1}^{d+1}\left( \theta _{\mu \nu }\right) &=&1
\end{eqnarray}
as far as eq(2.26) is concerned that is in the region of the
moduli space where the complex moduli $a_{i}$ are zero
$(i=1,\ldots )$. However, in the general case where the $a_{i}$'s
are non zero we should have moreover:
\begin{equation}
\prod_{\mu =1}^{d+1}\left( \theta _{\mu \nu }^{n_{\mu }^{\alpha }}\right)
=1;\quad \alpha =1,...,d+1.
\end{equation}
Let us comment briefly these constraint relations.  Eq(4.11)
reflects that the
parameters $\theta _{\nu \mu }$ are  just the inverse of $\theta _{\mu \nu }$%
\ and can be viewed as describing deformations away from the
identity suggesting by the occasion that they may be realized as
$$\theta _{\mu \nu }=\exp \eta _{\mu \nu }$$, where $\eta _{\mu
\nu }=-\eta _{\nu \mu }$ is the infinitesimal version of the non
commutativity parameter. The constraint (4.12-13) reflect just the
remarkable property according to which $U_{\nu }^{\frac{D}{\delta
_{\nu }}}$ and $\Pi _{\mu =1}^{d+2}\left( U_{\mu }\right) $ are
elements in the centre ${\cal{Z(A}}_{nc})$ of the non commutative
algebra $ {\cal{A}}_{nc},$i.e;
\begin{eqnarray}
\left[ U_{\mu },U_{\nu }^{\frac{D}{\delta _{\nu }}}\right] &=&0,
\\ \left[ U_{\mu },\Pi _{\nu =1}^{d+2}\left( U_{\nu }\right)
\right] &=&0.
\end{eqnarray}
Finally, the constraint eqs(4.14), obtained by requiring $\left[
U_{\mu },\Pi _{\nu =1}^{d+2}\left( U_{\nu }^{n_{\mu }^{\alpha
}}\right) \right] =0,$\ describe the compatibility between non
commutativity and deformations of the complex structure of the
Calabi-Yau hypersurfaces.\par In what follows we shall solve the
above constraint equations (4.11-14) in terms of toric geometry
data of the toric variety in which the mirror geometry is
embedded. Since these solutions depend on the weight vector
$\delta$ we will  consider  two cases; $%
\delta _{\mu }=1$ for all values of $\mu $   and $\delta _{\mu }$
taking general \ numbers eqs(2.20).
\subsubsection{ Matrix representations for  projective spaces  }
The analysis we have developed  so far can be made more explicit
by computing the NC  algebras associated to the Calabi-Yau
hypersurface orbifolds with discrete torsion.  In this regards,  a
simple and instructive  class of solutions of the above constraint
eqs may be worked in the framework of the $P^{d+1}$ ordinary
projective spaces. To do this, consider a  $d$ complex dimension
Calabi-Yau homogeneous hypersurfaces in $ P^{d+1}$ namely,
\begin{equation}
u_{1}^{d+2}+u_{2}^{d+2}+u_{3}^{d+2}+u_{4}^{d+2}+\ldots+u_{d+2}^{d+2}+a_{0}
\prod_{\mu =1}^{d+2}u_{\mu }=0,
\end{equation}
with the discrete isometries(2.31) and Calabi-Yau charges
$Q_{\mu}^{a}$  satisfying  \be
\sum_{\mu=1}^{d+2}Q_{\mu}^{a}=0,\qquad a=1,...,d.\ee From
constraint eq (4.12),  it is not difficult to see   that $\theta
_{\mu \nu }\ $ is an element of the $\bf Z_{d+2}$ group and so can
be written as
\begin{equation}
\theta _{\mu \nu }=\omega ^{L_{\mu \nu }},
\end{equation}
where $\omega= exp{2\pi i \over d+2}$ and where $L_{\mu \nu }$ is
a $(d+1)\times (d+1)$ antisymmetric matrix, i.e $  L_{\mu \nu }
=-L_{\nu \mu }$,  as required by eq(4.11). Putting this solution
back into eq(4.13), one discovers that this tensor should satisfy
\begin{equation}
\sum_{\mu =1}^{d+1}L_{\mu \nu } =0,\quad \mbox {modulo (d+2) }.
\end{equation}
Using the toric data of the Calabi-Yau manifold  $%
\sum_{\mu =1}^{d+1}Q_{\mu }^{a}=0$ and $\sum_{\mu =1}^{d+1}\xi
_{\mu }^{ \alpha }=0\mathbf{,}$ namely
\begin{eqnarray}
Q_{\mu } &=&\sum_{a=1}^{r}p_{a}\left( q_{\mu }^{a}+\frac{{1}%
}{d+1}\sum_{i=d+2}^{k}q_{i}^{a}\right), \\ \xi _{\mu }^{ \alpha }
&=&\sum_{a=1}^{r}p_{a}\left( q_{\mu
}^{a}n_{\mu }^{\alpha }+\frac{1}{d+1}\sum_{i=d+2}^{k+1}q_{i}^{a}n_{i}^{%
\alpha }\right),
\end{eqnarray}
one sees that the $L_{\mu \nu }$'s can be solved as bilinear forms of $%
Q_{\mu }^{ a}$ and $\xi _{\mu }^{\alpha }$ namely:
\begin{equation}
L_{\mu \nu }=L_{1}\Omega _{ab}Q_{\mu }^{ a}Q_{\nu }^{
b}+L_{2}\Omega _{\alpha \beta }\xi _{\mu }^{ \alpha }\xi _{\nu }^{
\beta }.
\end{equation}
 Here  $L_{1}$ and $L_{2}$\ are numbers modulo $(d+2)$ and
$\Omega _{ab} $\ and \ $\Omega _{\alpha \beta }$ are respectively
the antisymmetric $r\times r$\ and $(d+2)\times (d+2)$\ for even
integer values of $r$ and $d$  or their generalized expressions
otherwise.  Moreover, $L_{\mu \nu }$\ can also be rewritten in
terms of the ${\bf a}_{\mu }^{\nu }$
 components of \ $%
{\bf b}_{\mu }.$  For the particular case $L_2$=0, eq(4.23)
reduces to:
\begin{equation}
L_{\mu\nu}=-L_{\nu\mu}=m_{ab}Q_{\mu}^{[a}Q_{\nu}^{b]},
\end{equation}
where $m_{ab}$ is an antisymmetric $d\times d$ matrix of integers modulo ($%
d+2)$. It  satisfies
\begin{equation}
\sum_{\mu=1}^{d+2}L_{\mu\nu}=0.
\end{equation}
The NC  extension of eq(4.17) is given by the following algebra,
to which we refer to as ${\cal{A}}_{nc}(d+2);$
\begin{eqnarray}
U_{\mu}U_{\nu} &=&\omega _{\mu\nu}\varpi
_{\nu\mu}U_{\nu}U_{\mu};\qquad \ \mu,\nu=1,...,(d+1),   \nn\\
U_{\mu}U_{d+2} &=&U_{d+2}U_{\mu};\qquad \ \mu=1,...,(d+1),
\end{eqnarray}
where $\varpi _{\mu\nu}$ is the complex conjugate of $\omega
_{\mu\nu} $. The latters  are realized in terms of the Calabi-Yau
charges data as follows:
\begin{equation}
\omega _{\mu\nu}=\exp i\left( \frac{2\pi
}{d+2}m_{ab}Q_{\mu}^{a}Q_{\nu}^{b}\right) =\omega
^{m_{ab}Q_{\mu}^{a}Q_{\nu}^{b}}.
\end{equation}
Using the propriety $\varpi _{\mu\nu}^{d+2}=1$ and $\prod
\limits_\mu \varpi _{\mu\nu}=1$, one can check  that the center of
the algebra (4.26) is given by the
\be
{\cal{Z}}({\cal
A}_{nc})=\lambda_1U_1^{d+2}+\lambda_2U_2^{d+2}+\ldots+\lambda_{d+1}U_{d+1}^{d+2}
+\lambda_{d+2}I_{d+2}+\prod\limits_{\mu=1}^{d+1}U_\mu. \ee Schur
lemma implies that this matrix equation can be written
\be
{\cal{Z}}({\cal A}_{nc})=p(u_1,u_2,\ldots,u_{d+1})I_{d+2}.
 \ee
To determine the explicit expression of
$p(u_1,u_2,\ldots,u_{d+1})$, let us discuss in what follow the
matrix  irreducible representations of the NC Calabi-Yau
  algebra for a regular point. In the next subsection we will give the representation
   for  the  fixed points, where
   the representation  becomes reducible and corresponds to
   fractional branes.\\
 Finite dimensional representations of the algebra  (4.26) are
given by matrix subalgebras $Mat\left[ n(d+2),C\right] $, the
algebra of $n(d+2)\times n(d+2)$ complex matrices, with
$n=1,2,...$.  Computing the determinant of  both sides of
eqs(4.26)
\be
det\;(U_{\mu}U_{\nu}) =({\omega _{\mu\nu}\varpi _{\nu\mu}})^D
det\; (U_{\nu}U_{\mu})= det \;(U_{\nu}U_{\mu}),
 \ee
  the
dimension $D$ of the representation to be such that:
\begin{equation}
\left( \omega _{\mu\nu}\varpi _{\nu\mu}\right) ^{D}=1.
\end{equation}
Using the identity (4.19), one discovers that $D$ is a multiple of
$(d+2)$. We consider the fundamental $(d+2)\times (d+2)$ matrix
representation obtained by  introducing  the following set
$\left\{ {\mathbf{Q;P}}_{\alpha _{ab}};a,b=1,...,d\right\} $ of
matrices:
\begin{equation}
{\mathbf{P}}_{\alpha _{ab}}={diag(1,\alpha _{ab},}\alpha
_{ab}^{2},...,\alpha _{ab}^{d+1});\quad \mathbf{Q}=\left(
\begin{array}{ccccccc}
0 & 0 & 0 & . & . & . & 1 \\ 1 & 0 & 0 & . & . & . & 0 \\ 0 & 1 &
0 & . & . & . & 0 \\ . & . & . & . & . & . & . \\ . & . & . & . &
. & . & . \\ 0 & 0 & 0 & . & 1 & 0 & 0 \\ 0 & 0 & 0 & . & . & 1 &
0
\end{array}
\right)
\end{equation}
where $\alpha _{ab}= w^{m_{ab}} $ satisfying $\alpha
_{ab}^{d+2}=1$. From these expressions, it is not difficult to see
that the $\left\{ {\mathbf{Q;P}}_{\alpha
_{ab}};a,b=1,...,d\right\} $ matrices obey the algebra:
\begin{eqnarray}
{\mathbf{P}}_{\alpha }{\mathbf{P}} _{\beta
}&=&{\mathbf{P}}_{\alpha \beta }\nn\\ {\mathbf{P}}_{\alpha
}^{d+2}&=&1, \\
 {\mathbf{Q}}^{d+2}&=&1. \nn
\end{eqnarray}
Using the following identities
\begin{eqnarray}
  {\mathbf{P}}_{\alpha _{\mu}}^{n_{\mu}}{\mathbf{Q}}^{m_{\mu}} &=&\alpha
_{\mu}^{n_{\mu}m_{\mu}}{\mathbf{Q}}^{m_{\mu}}{\mathbf{P}}_{\alpha
_{\mu}}^{n_{\mu}},  \\ \left( {\mathbf{P}}_{\alpha
_{\mu}}^{n_{\mu}}{\mathbf{Q}}^{m_{\mu}}\right) \left(
{\mathbf{P}}_{\alpha
_{\nu}}^{n_{\nu}}{\mathbf{Q}}^{m_{\nu}}\right)
&=&\alpha _{\mu}^{n_{i}m_{\nu}}\alpha _{\nu}^{-m_{\mu}%
n_{\nu}}\left( {\mathbf{P}}_{\alpha _{\nu}}^{n_{\nu}}{\mathbf{Q}}%
^{m_{\nu}}\right) \left( {\mathbf{P}}_{\alpha _{\mu}}^{n_{\nu}}{\mathbf{Q}}%
^{m_{\mu}}\right) ,
\end{eqnarray}
one can check that the $U_\mu$ operators can be  realized as
\begin{equation}
U_{\mu}=u_{\mu}\prod_{a,b=1}^{d}\left( {\mathbf{P}}_{\alpha
_{ab}}^{Q_{\mu}^{a}}{\mathbf{Q}}^{Q{\mu}^{b}}\right),
\end{equation}
where $u_{\mu}$ are $C$-number which should be thought of as in
(4.17).  From the   Calabi-Yau condition, one can also check that
the above representation satisfy
\begin{eqnarray}
U_{\mu}^{d+2}&=&
u_{\mu}^{d+2}{\mathbf{I}}_{d+2}\nn\\\prod_{\mu=1}^{d+1}U_{\mu}
&=&{\mathbf{I}}_{d+2}\left( \prod_{\mu=1}^{d+1}u_{\mu}\right).
\end{eqnarray}
Putting these relations  back into (4.29), one finds that the
polynomial $p(u_\mu)$ is nothing but  the eq (4.17) of the
Calabi-Yau hypersurface.

\subsubsection{ Solution for weighted  projective spaces}

In the case of weighted projective spaces with a weight vector ${%
\delta =}(\delta _{1},...,\delta _{d+2})$; the degree $D$ of the
Calabi-Yau polynomials and the corresponding ${N}_{i}$\ vertices
are respectively given by eqs(2.19-20) and (2.24-25). Note that
integrality of the vertex entries requires that $D$ should be the
smallest commun multiple of the weights $\delta _{\mu };$ that is
$\frac{D}{\delta _{\mu }}$ an integer.\ Following the same
reasoning as for the case of the projective space, one can work
out a class of solutions of the constraint eqs(4.11-13) in terms
of powers of $\omega _{_{\mu }}$. We get the result
\begin{equation}
\theta _{\mu \nu }=\exp i2\pi \left[ \frac{(\delta _{\nu })L_{\mu \nu }}{D}%
\right] ,
\end{equation}
where $L_{\mu \nu }$\ is as in eq(4.23). Instead of  being
general, let  us consider  a concrete example dealing with the
analogue of the quintic
 in the weighted projective space WP$_{\{\delta _{1},\delta
_{2},\delta _{3},\delta _{4},\delta _{5}\}}^{4}$. In this case the
Calabi-Yau  hypersurface, $ \sum_{\mu =1}^{5}u_{\mu
}^{\frac{D}{\delta _{\mu }}}+a_{0}\prod_{\mu =1}^{5}\left( u_{\mu
}\right) =0$; which for the example $\delta _{1}=2$ and $\delta
_{2}=\delta _{3}=\delta _{4}=\delta _{5}=1$, reduces to:
\begin{equation}
u_{1}^{3}+u_{2}^{6}+u_{3}^{6}+u_{4}^{6}+u_{5}^{6}+a_{0}\prod_{\mu
=1}^{5}\left( u_{\mu }\right) =0.
\end{equation}
This polynomial has discrete isometries acting on the homogeneous
coordinates $u_{\mu }$ as:
\begin{equation}
u_{\mu }\rightarrow u_{\mu }\zeta _{\mu }^{{\bf a}_{\mu
}^{\nu}}\qquad \mu =1,...,5 ,
\end{equation}
with $\zeta _{1}^{3}=1$ while $\zeta _{\mu}^{6}=\omega ^{6}=1;$
i.e $\zeta _{1}=\omega ^{2}$ and $\zeta _{\mu}=\omega  $ and where
the ${\bf a}_{\mu }^{\nu}$ are consistent with  the Calabi-Yau
condition. In the coordinate patch $\left\{ u_{\mu}\right\}
_{1\leq 4}$ with $u_{5}=1$, the  equations defining  the NC
geometry of the Calabi-Yau (4.39) with discrete torsion, upon
using the correspondence $u \to U$, are given by the algebra (5.1)
where the $\theta _{\mu\nu}$ parameters should obey now the
following constraint eqs:
\begin{eqnarray}
\theta _{\mu1}^{3} &=&1, \quad \mu=2,3,4,\nn \\ \theta
_{\mu\nu}^{6} &=&1,\quad \nu\neq 1,\mu \\ \prod_{\mu=1}^{4}\theta
_{\nu\mu} &=&1,\qquad \forall \nu \nn\\ \theta _{\mu\nu}\theta
_{\nu\mu} &=&1,\qquad \forall \mu,\nu\nn.
\end{eqnarray}
Setting $\theta _{\mu\nu}$ as $\theta _{\mu\nu} =\omega
^{L_{\mu\nu}}$ the constraints on $L_{\mu\nu}$ read as:
\begin{eqnarray}
 L_{\mu\nu}&=&-L_{\nu\mu} \ \ \ \mbox{integers  modulo 6}, \nn  \\
L_{\mu 1} &=& \mbox{even  modulo 6}.
\end{eqnarray}
Particular solutions of this geometry   may be  obtained by
using antisymmetric bilinears of ${\bf a}_{\mu }^{\nu}$.
Straightforward calculations show that, for $p_\mu=1$,
$L_{\mu\nu}$ is given by the following $4\times 4$ matrix:
\begin{equation}
L_{\mu\nu}=\left(
\begin{array}{cccc}
0 & k_{1}-k_{3} & -k_{1}+k_{2} & k_{3}-k_{2} \\ -k_{1}+k_{3} & 0 &
k_{1} & -k_{3} \\ k_{1}-k_{2} & -k_{1} & 0 & k_{2} \\ -k_{3}+k_{2}
& k_{3} & -k_{2} & 0
\end{array}
\right)
\end{equation}
where the $k_{\mu}$ integers are such that $ k_{\mu}-k_{\nu}\equiv
2r_{\mu\nu}\in 2Z.$ \\ The NC  algebra associated with eq(4.39)
reads, in terms of $\omega _{\mu}=\omega ^{k_{\mu}}$ and $\varpi
_{\mu}=\omega ^{-k_{\mu}} ,$ as follows:
\begin{eqnarray}
U_{1}U_{2} &=&\omega _{1}\varpi _{3}U_{2}U_{1},\qquad
U_{1}U_{3}=\varpi _{1}\omega _{2}U_{3}U_{1}, \nn \\ U_{1}U_{4}
&=&\omega _{3}\varpi _{2}U_{4}U_{1},\qquad U_{2}U_{3}=\omega
_{1}U_{3}U_{2}, \\
U_{2}U_{4} &=&\varpi _{3}U_{4}U_{2},\qquad %
U_{3}U_{4}=\omega _{2}U_{4}U_{3}\nn.
\end{eqnarray}
Furthermore taking $\ \alpha =\omega _{1}\varpi _{3},$ $\beta
=\omega _{2}\varpi _{3}$ and $\gamma =\omega _{3}$, one discovers
an extension of the $BL$ NC  algebra( 4.4); the difference is that
now the deformation parameters are such that:
\begin{equation}
\alpha ^{3}=\beta ^{3}=\gamma ^{6}=1.
\end{equation}

\subsubsection{Fractional Branes  }
 Here  we study  the fractional branes corresponding  to reducible
 representations at singular points.  To illustrate the idea,  we  give a concrete example
concerning   the mirror
  geometry in terms
   of the ${\bf P}^{d+1}$ projective space.  First  note  that
    the ${\cal{A}}_{nc}(d+2)$ (4.37)
 corresponds to regular points
of NC  Calabi-Yau. This  solution  is  irreducible and the branes
do not fractionate. A similar solutions may be worked out as well
for fixed  points   where we  have   fractional branes.  We  focus
our attention on the  orbifold of the {\it{eight-tic}},  namely,
\begin{equation}
u_{1}^{8}+u_{2}^{8}+\ldots+u_{8}^{8}+a_{0} \prod_{\mu
=1}^{8}u_{\mu }=0,
\end{equation}
with the discrete isometries ${\bf {Z}}_{8}^{6}$ and Calabi-Yau
charges  $Q_{\mu}^{a}$
\begin{eqnarray}
Q_{\mu}^{1} &=&(1,-1,0,0,0,0,0,0)  \nn \\ Q_{\mu}^{2}
&=&(1,0,-1,0,0,0,0,0) \nn\\ Q_{\mu}^{3} &=&(1,0,0,-1,0,0,0,0)\nn\\
Q_{\mu}^{4} &=&(1,0,0,0,-1,0,0,0)\\
 Q_{\mu}^{5} &=&(1,0,0,0,0,-1,0,0)\nn\\
 Q_{\mu}^{6} &=&(1,0,0,0,0,0,-1,0).\nn
\end{eqnarray}
 The corresponding NC  algebra is deduced from  the general  one given in (4.26).
   At  regular points, the matrix theory representation of this
   algebra  is irreducible as shown on eqs (4.37).
 However, the situation is more subtle at  fixed points  where representations are reducible. One way to
deal with the singularity of the orbifold with respect to ${\bf
{Z}}_{8}^{6}$ is to interpret the algebra as describing a ${\bf
Z}_{8}^{3}$ orbifold with ${\bf Z}_{8}^{3}$\ discrete torsions
having singularities in codimension four. Starting from eqs(4.26)
and choosing matrix coordinates $U_{5}$, $U_{6}$ and $U_{7}$ in
the centre of the algebra by setting
\begin{equation}
\left( \omega _{\mu\nu}\varpi _{\nu\mu}\right) =1,\quad  \mbox{
for}\;\;\mu=5,6,7,8;\qquad \forall \nu=1,...,8,
\end{equation}
the algebra reduces to \bea
 U_{1}U_{2} &=&\alpha _{1}\alpha
_{2}U_{2}U_{1}\nn\\ U_{1}U_{3}&=&\alpha _{1}^{-1}\alpha
_{3}U_{3}U_{1}\nn\\ U_{1}U_{4} &=&\alpha _{2}^{-1}\alpha
_{3}^{-1}U_{4}U_{1}\\
 U_{2}U_{3}&=&\alpha _{1}U_{3}U_{2}\nn\\
 U_{2}U_{4}
&=&\alpha _{2}U_{4}U_{2},\nn\\ U_{3}U_{4}&=&\alpha
_{3}U_{4}U_{3}\nn \eea
 and all remaining other relations are commuting. In these
equations, the $\alpha _{\mu}$ 's are such that $\alpha
_{\mu}^{8}=1;$ these are the phases  ${\bf Z}_{8}^{3}$.  At  the
singularity where the $u_{1}$, $u_{2},$ $u_{3},$ and $u_{4}$
moduli of eq(4.37) go to zero, one ends with the familiar result
for orbifolds with discrete torsion. Therefore the D-branes
fractionate in the codimension four singularities of the eight-tic
geometry.

\section{Link with the BL Construction}
In this section  we want to rederive the results of \cite{39}
concerning NC  quintic using the analysis developed in section 3
and 4. Recall that in the coordinate patch $\left\{
u_{\mu}\right\} _{1\leq 4}$ and $u_{5}=1$, the defining equations
of NC  geometry of the quintic with discrete torsion, upon using
the correspondence $u \to U$, are given by the following operators
algebra.
\begin{equation}
U_{\mu}U_{\nu}=\theta _{\mu\nu}U_{\nu}U_{\mu},\qquad
\nu>\mu=1,...,4,
\end{equation}
where the $\theta _{\mu\nu}$'s are non zero complex parameters.
 As the monomials $U_{\mu}^{5}$ and
$
\prod_{\mu=1}^{5}\left( U_{\mu}\right) $ are  commuting
 with all the $%
U_{\mu}$ 's , we have also
\begin{eqnarray}
\left[ U_{\nu},U_{\mu}^{5}\right] &=&0, \nn  \\ \left[
U_{\nu},\prod_{\mu=1}^{4}U_{\mu}\right] &=&0.
\end{eqnarray}
 Compatibility between eqs (5.1-2) gives constraint relations on $\theta _{\mu\nu}$'s
 namely
\begin{eqnarray}
\theta _{\nu\mu}^{5} &=&1, \\ \prod_{\mu=1}^{4}\theta _{\nu\mu}
&=&1,\qquad \forall \nu \\ \theta _{\mu\nu}\theta _{\nu\mu}
&=&1;\qquad \theta _{\mu5}=1,\qquad \forall \mu,\nu.
\end{eqnarray}
To establish the link between our way of doing and the
construction of \cite{41}, it is interesting to note that the
analysis of \cite{41} corresponds in fact to a special
representation of the formalism we developed so far. The idea is
summarized  as follows: First start from eq(3.1), which reads for
the quintic as:
\begin{equation}
u_{\mu }\rightarrow u_{\mu }\omega ^{{\bf b}_{\mu }},
\end{equation}
where the ${\bf b}_{\mu }$ weights, ${\bf b}_{\mu
}=\sum_{\nu=1}^{5}{\bf a}_{\mu }^{\nu}, \mu=1,\ldots,5$, are such
that
 \bea
 \sum\limits_{\nu=1}^5{\bf b}_{\mu }&=&0.
 \eea
This relation, interpreted as the Calabi-Yau condition, can be
 solved in different ways. A way to do is to set the
 ${\bf b}_{\mu }$ weights as  \be
{\bf b}_{\mu }=(p_{1}+p_{2}+p_{3},-p_{1},-p_{2},-p_{3},0), \ee or
equivalently by taking the weight components ${\bf b}_{\mu
}^{\nu}$  as:
 \be
{\bf a}_{\;\;\mu}^{\nu} =\left(
\begin{array}{ccccccc}
p_1 & p_2 & p_3 & 0 & 0  \\ -p_1 & 0 & 0 & 0 & 0 \\ 0 & -p_2 & 0 &
0 & 0  \\ 0 & 0 & -p_3 & 0 &0\\0 & 0 & 0 & 0 &  0
\end{array}
\right), \ee where $p_{a}$ are integers modulo $5$. More general
solutions can be read from eqs (4.23) by following the same
method. The next step is to take $\theta_{\mu\nu}=exp
i({2\pi\over5}L_{\mu\nu})$ with $L_{\mu\nu}$ as follows:
\begin{equation}
L_{\mu\nu}=m_{12}\left( {\bf a}_{\mu}^{1}{\bf a}_{\nu}^{2}-{\bf
a}_{\nu}^{1}{\bf a}_{\mu}^{2}\right)
-m_{23}\left( {\bf a}_{\mu}^{2}{\bf a}_{\nu}^{3}-{\bf a}_{\nu}^{2}{\bf a}_{\mu}^{3}\right)
+m_{13}%
\left( {\bf a}_{\mu}^{1}{\bf a}_{\nu}^{3}-{\bf a}_{\nu}^{1}{\bf
a}_{\mu}^{3}\right),
\end{equation}
where $m_{12}=k_{1}$, $m_{23}=k_{2}$ and $m_{13}=k_{3}$ are
integers modulo $5$. For $p_\mu=1$, we get
\begin{equation}
L_{\mu\nu}=\left(
\begin{array}{cccc}
0 & k_{1}-k_{3} & -k_{1}+k_{2} & k_{3}-k_{2} \\ -k_{1}+k_{3} & 0 &
k_{1} & -k_{3} \\ k_{1}-k_{2} & -k_{1} & 0 & k_{2} \\ -k_{3}+k_{2}
& k_{3} & -k_{2} & 0
\end{array}
\right),
\end{equation}
and so the NC  quintic algebra reads:
\begin{eqnarray}
U_{1}U_{2} &=&\omega ^{k_{1}-k_{3}}U_{2}U_{1},\qquad
U_{1}U_{3}=\omega ^{-k_{1}+k_{2}}U_{3}U_{1},  \nn\\ U_{1}U_{4}
&=&\omega ^{k_{3}-k_{2}}U_{4}U_{1},\qquad U_{2}U_{3}=\omega
^{k_{1}}U_{3}U_{2}, \\ U_{2}U_{4} &=&\omega
^{-k_{3}}U_{4}U_{2},\qquad U_{3}U_{4}=\omega
^{k_{2}}U_{4}U_{3}.\nn
\end{eqnarray}
Setting $\omega _{\mu}=\omega ^{k_{\mu}}$ and $\varpi
_{\mu}=\omega ^{-k_{\mu}}\ ,$\ the above relations become:
\begin{eqnarray}
U_{1}U_{2} &=&\omega _{1}\varpi _{3}U_{2}U_{1},\qquad
U_{1}U_{3}=\varpi _{1}\omega _{2}U_{3}U_{1},  \nn\\ U_{1}U_{4}
&=&\omega _{3}\varpi _{2}U_{4}U_{1},\qquad U_{2}U_{3}=\omega
_{1}U_{3}U_{2}, \\
U_{2}U_{4} &=&\varpi _{3}U_{4}U_{2},\qquad U_{3}U_{4}=\omega _{2}%
U_{4}U_{3}\nn.
\end{eqnarray}
Now taking $\alpha =\omega _{1}\varpi _{3},$ $\beta =\omega
_{2}\varpi _{3}$ and $\gamma =\omega _{3}$, one discovers exactly
the $BL$ algebra eqs(4.8).
\subsection{More on the NC  quintic }
As we mentioned, the solution given by eqs(5.8-9) is in fact a
special realization of the BL algebra (4.8). One can also write
down other representations of the NC  quintic; one of them is
based on taking ${\bf a}_{\mu }^{\nu}$ as:
\begin{equation}
{\bf a}_{\;\;\mu}^{\nu}=\left(
\begin{array}{ccccccc}
p_1 & 0 & p_3 & 0 & 0  \\ -2p_1 & p_2 & 0 & 0 & 0 \\ p_1  & -2p_2
& p_3 & 0 & 0
\\ 0 & p_2 & -2p_3 & 0 &0\\0 & 0 & 0 & 0 & 0
\end{array}
\right).
\end{equation}
The corresponding ${\bf b}_{\mu }$  weight vector is then:
\begin{equation}
{\bf b}_{\mu }=(p_{1}+p_{3},-2p_{1}+p_{2},p_{1}-2p_{2}+p_{3},
p_{2}-2p_{3};0) ,
\end{equation}
with $p_{a}$'s are integers modulo $5$. As one sees this is a
different solution from that given in eq(5.8-9) as
 the corresponding $\Gamma $ group of isometries acts differently on the
$u_{\mu }$ variables leading then to a different orbifold with
discrete torsion. Note that setting $p_{\mu}=1$,
  the ${\bf a}_\mu^\nu$ weights are nothing but the Mori vectors of the blow up of the
  the $\hat A_2$ affine singularity of K3, used in
the geometric engineering method of  $4D$ $ {\cal N}=2$
superconformal theories embedded in type II superstrings.
\\ Setting $p_\mu=1$ and using equations (5.10) and (5.14), the
anti-symmetric $ L_{\mu\nu}$ matrix reads as:
\begin{equation}
L_{\mu\nu}=\left(
\begin{array}{cccc}
0 & k_{1}+k_{2}+2k_{3} & -2k_{1}-2k_{2} & k_{1}+k_{2}-2k_{3} \\
-k_{1}-k_{2}-2k_{3} & 0 & 3k_{1}-k_{2}-2k_{3} &
-2k_{1}+2k_{2}+4k_{3} \\ 2k_{1}+2k_{2} & -3k_{1}+k_{2}+2k_{3} & 0
& k_{1}-3k_{2}-2k_{3} \\ -k_{1}-k_{2}+2k_{3} &
2k_{1}-2k_{2}-4k_{3} & -k_{1}+3k_{2}+2k_{3} & 0
\end{array}
\right)
\end{equation}
where the $k_{1},k_{2}$ and $k_{3}$ are integers modulo $5$. The
new algebra describing the NC  quintic reads, in terms of the
$\omega _{\mu}$ and $\varpi _{\nu}$ generators of the $Z_{5}^{3}$,
as follows:
\begin{eqnarray}
U_{1}U_{2} &=&\omega _{1}\omega _{2}\omega
_{3}^{2}U_{2}U_{1},\qquad U_{1}U_{3}=\varpi _{1}^{2}\varpi
_{2}^{2}U_{3}U_{1},\nn \\ U_{1}U_{4} &=&\omega _{1}\omega
_{2}\varpi _{3}^{2}U_{4}U_{1},\qquad U_{2}U_{3}=\omega
_{1}^{3}\varpi _{2}\varpi _{3}^{2}U_{3}U_{2}, \\
U_{2}U_{4} &=&\varpi _{1}^{2}\omega _{2}^{2}\omega _{3}^{4}%
U_{4}U_{2},\qquad U_{3}U_{4}=\omega _{1}\varpi _{2}^{3}\varpi
_{3}^{2}U_{4}U_{3}\nn.
\end{eqnarray}
Setting $\alpha =\omega _{1}\omega _{2}\omega _{3}^{2},$ $\beta
=\varpi _{1}\varpi _{2}\omega _{3}^{2}$ and $\gamma =\omega
_{1}^{2}\varpi _{2}^{2}\varpi _{3}^{4}$, one discovers, once
again,  the $BL$ algebra (4.7). Therefore eq(5.9) and eq(5.14)
give two representations of the BL algebra.
\subsection{ Comments on lower dimension  CY manifolds}
The analysis we developed so far applies to complex $d$ dimension
homogeneous hypersurfaces with discrete torsion; $d\geq 2$. We
have discussed the cases $d\geq 3$; here we want to complete this
study for lower dimension Calabi-Yau manifolds namely $K3$ and the
elliptic curve. These are very special cases which deserves some
comments. For the $K3$    surface in $CP^3$, we have
\begin{equation}
u_{1}^{4}+u_{2}^{4}+u_{3}^{4}+u_{3}^{4}+a_{0}\prod_{\mu=1}^{4}u_{\mu}=0,
\end{equation}
This is a quartic polynomial with a $\bf Z_{4}\t Z_4$ symmetry
acting on the $u_i$ variables as:
\be
u_\mu\to w^{Q_\mu^a}u_\mu,\ee where $w^4=1$ and  where  ${\bf
{a}}_\mu^a$ are integers satisfying the  Calabi-Yau condition
$\sum\limits_{\mu=1}^4 Q_\mu^a=0$. Choosing $Q_\mu^a$ as, \bea
Q_\mu^1=(1,-1,0,0)\nn\\ Q_\mu^2=(1,0,-1,0)
 \eea
 the $3\times3$ matrix $L_{\mu\nu}$ reads as
\begin{equation} L_{\mu\nu}=\left(
\begin{array}{cccc}
0 & k & -k \\ -k & 0 & k
 \\ k & -k & 0
\end{array}
\right).
\end{equation}
Therefore the NC  $K3$ algebra reads as:
\begin{eqnarray}
U_{1}U_{2} &=&U_{2}U_{1}e^{i\frac{2\pi k}{4}},  \nn \\ U_{1}U_{3}
&=&U_{3}U_{1}e^{-i\frac{2\pi k}{4}}, \nn\\ U_{1}U_{4}
&=&U_{4}U_{1},
\\ U_{2}U_{3} &=&U_{3}U_{2}e^{i\frac{2\pi k}{4}} ,  \nn \\
U_{2}U_{4} &=&U_{4}U_{2}, \nn  \\ U_{3}U_{4} &=&U_{4}U_{3 }.\nn
\end{eqnarray}
where $k$ is an integer modulo 4. Note that one gets similar
results by making other choices of  $Q_i^a$ such as, \bea
Q_\mu^1=(1,-2,1,0)\nn\\ Q_\mu^2=(1,1,-2,0).
 \eea
 More general results may also be written down for $K3$ embedded in
  $WP_{(\delta_1,\delta_2,\delta_3,\delta_4)}$.
  In the case of the one dimensional
elliptic fiber given by a cubic in $P^{2}$ \begin{equation}
u_{1}^{3}+u_{2}^{3}+u_{3}^{3}+a_{0}\prod_{\mu=1}^{3}u_{\mu}=0,
\end{equation}
the constraint equations defining non commutativity are trivially
solved. They show that $L_{\mu\nu}=0$ and so $\theta_{12}=1$
leading then to a commutative geometry. NC  geometries involving
elliptic curves can be constructed; the idea is to consider
orbifolds of products of elliptic curves. More details are exposed
in the following section. Related ideas  with fractional branes
will be considered as well.

\section{NC Elliptic  Manifolds}

In this section we want refine the study  of  the NC  Calabi-Yau
hypersurface defined in terms of orbifolds of elliptic curves. The
original idea of this construction was introduced first in
\cite{39}; see also \cite{73}, in connection with the NC orbifold
$\frac{T^{6}}{Z_{2}^{2}}.$ The method is quite similar to that
discussed for the quintic and generalized
 Calabi-Yau geometries  in sections 4 and
5. To start consider the following elliptic realization of
 $\frac{T^{2n+2}}{%
\Gamma }$; that is $T^{2n+2}$ is represented by the product of
$(n+1)$ elliptic curves $\left( T^{2}\right) ^{\otimes (2k+1)}$
where $n=2k$. Each  elliptic curve is given in Weierstrass form
as:
\begin{equation}
y_{\mu }^{2}=x_{\mu }(x_{\mu }-1)(x_{\mu }-a_{\mu }),\quad \mu =1,...,n+1,
\end{equation}
with a point added at infinity $\mu =1,...,n+1$. The system
$\{(x_{\mu },y_{\mu });\quad \mu
=1,...,n+1\}$ defines the complex coordinates of $ C^{2n+2}$ space and $%
a_{\mu }$ are $(n+1)$ complex moduli. For later use, we introduce
the algebra ${\cal{A}}_{c}$ of holomorphic functions on $
T^{2n+2}$. This is a commutative algebra generated by monomials in
the $x_{\mu }$'s and $ y_{\mu }$'s with the conditions eqs(6.1).
The discrete group $\Gamma $ acts on $x_{\mu }$'s and $y_{\mu }$'s
as:
\begin{eqnarray}
x_{\mu } &\rightarrow &x_{\mu }^{\prime }=x_{\mu },\nn \\ y_{\mu }
&\rightarrow &y_{\mu }^{\prime }=y_{\mu }\omega ^{Q_{\mu }},
\end{eqnarray}
where $\omega $ is an element of the discrete group $\Gamma $ and where $%
Q_{\mu }$\ are integers which should be compared with eq(4.24).
Note that if
one is requiring that eqs(6.1) to be invariant under ${\Gamma }$, then%
$\ \omega ^{2}$ should be equal to one that is $\omega =\pm 1.$ If one
requires moreover that the monomial $\prod_{\mu =1}^{n+1}{y_{\mu }}$ or
again the holomorphic $((n+1),0)$ form $d{y}_{1}\wedge d{y}_{2}....\wedge d{y}%
_{n+1},$ to be invariants under the orbifold action, it follows then that $%
\prod_{\mu =1}^{n+1}\omega ^{Q_{\mu }}=\omega ^{\Sigma _{\mu
}Q_{\mu }}=1$. This is equivalent to
\begin{equation}
\sum_{\mu =1}^{n+1}Q_{\mu }=0,\quad \mbox {modulo 2}
\end{equation}
defines the Calabi-Yau condition for the orbifold
$\frac{T^{2n+2}}{\Gamma }.$ Therefore the $\Gamma $\ discrete
group is given by ${\Gamma =}\left( \mathbf{{Z_{2}}}\right)
^{\otimes n}$. Following the discussion we made in section 4, this
equation can also be rewritten as
\begin{equation}
\sum_{\mu =1}^{n+1}Q_{\mu }^{a}=0,\quad \mbox{ modulo 2};\quad
a=1,...,n.
\end{equation}
The four fixed points of the orbifold for each two torus $T^{2}$
are located
at $(x_{\mu }=0,1,a_{\mu };$ $y_{\mu }=0)$ and the point at infinity; i.e $%
(x_{\mu }=\infty ;$ $y_{\mu }=\infty )$. This later can be brought
to a fixed finite point by working in another coordinate patch
related to the old one by using the change of variables:
\begin{eqnarray}
\ y_{\mu } &\rightarrow &y_{\mu }^{\prime }=\frac{y_{\mu }}{x_{\mu
}^{2}}\nn \\ x_{\mu } &\rightarrow &x_{\mu }^{\prime
}=\frac{1}{x_{\mu }}.
\end{eqnarray}
The NC  version of the orbifold $\frac{T^{2n+2}}{\Gamma }$ is
obtained by substituting the usual commuting $x_{\mu }$\ and
$y_{\mu }$ variables by the matrix operators $X_{\mu }$ and
$Y_{\mu }$\ respectively. These matrix operators satisfy the
folowing NC algebra structure:
\begin{eqnarray}
Y_{\mu }Y_{\nu } &=&\theta _{\mu \nu }Y_{\nu }Y_{\mu }, \\
X_{\mu }X_{\nu } &=&X_{\nu }X_{\mu }, \\
X_{\mu }Y_{\nu } &=&Y_{\nu }X_{\mu },
\end{eqnarray}
with
\begin{equation}
Y_{\mu }Y_{\nu }^{2}=Y_{\nu }^{2}Y_{\mu },
\end{equation}
as it is required by eq(6.1) and
\begin{equation}
\left[ Y_{\mu },\prod_{\nu =1}^{n+1}Y{_{\nu }}\right] =0.
\end{equation}
Like for the case of the  homogeneous hypersurfaces we considered
in sections 4 and 5,
here also the Calabi-Yau condition is fulfilled by imposing that the $%
\prod_{\nu =1}^{n+1}{Y_{\nu }}$ belongs to the centre of the NC algebra \ $%
{\cal{A}}_{nc}.$ Now using eqs(6.6-10), one gets the explicit
expression of the $\theta _{\mu \nu }$'s by solving the following
constraint eqs:
\begin{eqnarray}
\theta _{\mu \nu }\theta _{\mu \nu } &=&1, \\
\prod_{\mu =1}^{n+1}{\theta _{\mu \nu }} &=&1, \\
\theta _{\mu \nu }\theta _{\nu \mu } &=&1,\quad \theta _{\mu \mu }=1.
\end{eqnarray}
\ Note that eq(6.11) is a strong constraint which will have a
drastic consequence on the solving of noncommutativity. Comparing
this relation to eq(6.12), one can write
\begin{eqnarray}
\theta _{\mu \nu } &=&(-1)^{L_{\mu \nu }}, \nn\\ \sum_{\mu
=1}^{n+1}L_{\mu \nu } &=&0,\quad \mbox{modulo 2},
\end{eqnarray}
where $L_{\mu \nu }$ is  antisymmetric matrix, $L_{\mu \nu
}=-L_{\nu \mu }$, of integer entries given by
\begin{equation}
L_{\mu \nu }=\Omega _{ab}Q_{\mu }^{a}Q_{\nu }^{b};
\end{equation}
where $\Omega _{ab}=-\Omega _{ba},$ and $\Omega _{ab}=1$ for \
$a<b$.  This relation should be compared to eq(4.25). Moreover,
one learns from eq(6.14) that two cases should be distinguished.
The first one corresponds to the case $\theta _{\mu \nu }=-1$
$\forall $ $\mu \neq \nu ;$ that is;
\begin{equation}
L_{\mu \nu }=1;\quad \, \mbox{ modulo 2}.
\end{equation}
In this case, the constrained (6.12) is fulfilled provided $n$ is even; i.e $%
n=2k.$ So the group $\Gamma $ is given by ${\Gamma =}\left( \mathbf{{%
Z_{2}}}\right) ^{\otimes 2k}$. The second case corresponds to the
situation where some $\theta _{\mu \nu }$'s are equal to one:
\begin{eqnarray}
L_{\mu \nu } &=&1;\quad \mbox{ modulo 2}\quad \mu
=1,...,(n+1-r);\quad \mu \neq \nu   \\ L_{\mu \nu } &=&0;\quad
\mbox{ modulo 2};\quad \mu =(n-r+2),...,n+1,
\end{eqnarray}
where we have rearranged the variables so that the  matrix takes
the form
\begin{equation}
L_{\mu \nu }=\left(
\begin{array}{cc}
L_{\mu' \nu' }^{\prime } & \mathbf{0} \\ \mathbf{0} & \mathbf{0}
\end{array}
\right)
\end{equation}
In this case eq(6.17) shows that $n$ is even if $r$ is even and
odd if $r$ is odd. In what follows we build the solutions of the
NC  algebra(6.6-8) using finite dimensional matrices.
\subsection{Solution I}
Replacing the relation (6.16) back into\ in eqs(6.6-8), the non
commutativity algebra, which reads as:
\begin{eqnarray}
Y_{\mu }Y_{\nu } &=&-Y_{\nu }Y_{\mu },  \\ Y_{\mu }Y_{\nu }^{2}
&=&Y_{\nu }^{2}Y_{\mu }   \\ X_{\mu }X_{\nu } &=&X_{\nu }X_{\mu },
\\ X_{\mu }Y_{\nu } &=&Y_{\nu }X_{\mu },
\end{eqnarray}
may be realized in terms of $2^{k}\times 2^{k}$\ matrices of the
space of matrices $M(2^{k},C)$. These are typical relations
naturally solved by using the 2k dimensional Clifford algebra
generated by the basis system $\{\Gamma ^{\imath
},\mu=1,...,2k\}$:
\begin{eqnarray}
\{\Gamma ^{\mu},\Gamma ^{\nu}\} &=&2\delta^{\mu\nu},  \nn\\
\{\Gamma ^{i},\Gamma ^{2k+1}\} &=&0,
\end{eqnarray}
where $\Gamma ^{2k+1}=\prod_{i=1}^{2k}{\Gamma ^{i}}$. We therefore have:
\begin{eqnarray}
Y_{\mu} &=&b_{\mu}\Gamma ^{\mu},\quad \mu=1,...,2k,   \\ Y_{2k+1}
&=&b_{0}\Gamma ^{2k+1}, \\ X_{\mu } &=&x_{\mu }I_{2^{k}}
\end{eqnarray}
where the $b_{\mu }$'s are complex scalars.   This solution has
remarkable features: (i) After choosing a hermitian $\Gamma$
matrices representation,  one can see  at the fixed planes,  where
$2k$ variables among the (2k+1)
  $y_\mu$'s act by zero and all others zero, that  there exists a multiplicity of
   inequivalent representations for each set of roots $x_\mu$ of the Weierstrass forms.
   Therefore,  one can get $2^{k}$ distinct  NC  points,
as there are $2^{k}$ irreducibles representations corresponding to
$2^{k}$ eigenvalues of the non zero  matrix variable and so the
branes fractionate  on  the singularity. (ii) The non-commutative
points of the singular planes are then seen to be a $2k$  cover of
the commutative singular plane, which is a $(CP^1)^{\otimes k}$.
 The $2k$  cover is branched around the four points $x_k=0,1,a_k,\infty$ and hence
 the NC points form an elliptic manifold $T^{2k}$ of the form eq(6.1).
 Around each of these four points, there is a $(\bf Z_2)$ monodromy of the representations,
  which is characteristic of the local singularity as measuring the effect of discrete torsion.
\subsection{Solution II}

Putting the relations (6.17) back into\ in eqs(6.6-8), the
resulting NC  algebra depends  on the integer $r$ and reads
as:\newline
\begin{eqnarray}
Y_{\mu }Y_{\nu } &=&-Y_{\nu }Y_{\mu },\quad \mu ,\nu
=1,...,(n+1-r).
\\
Y_{\mu }Y_{\nu } &=&Y_{\nu }Y_{\mu },\quad \mu ,\nu =(n+2-r),...,(n+1),
 \\
Y_{\mu }Y_{\nu }^{2} &=&Y_{\nu }^{2}Y_{\mu },\quad \mu
=1,...,(n+1), \\ X_{\mu }X_{\nu } &=&X_{\nu }X_{\mu },   \\ X_{\mu
}Y_{\nu } &=&Y_{\nu }X_{\mu }.
\end{eqnarray}
For $r=2s$ even, irreducible representations of this algebra are
given, in terms of $2^{k-s}\times 2^{k-s}$\ matrices of the space
$M(2^{k-s},C),$ by the $2(k-s)$ dimensional Clifford algebra. The
result is:
\begin{eqnarray}
Y_{\mu} &=&b_{\mu}\Gamma ^{\mu},\quad i=1,...,2(k-s),   \\
Y_{2(k-s)+1} &=&b_{0}\prod_{\mu=1}^{2(k-s)}{\Gamma ^{\mu}}, \\
Y_{\mu} &=&y_{\mu}I_{2^{k-s}},i=2(k-s+1),...,(2k+1),   \\ X_{\mu }
&=&x_{\mu }I_{2^{k-s}}.
\end{eqnarray}
In the end of this section, we would like to note that this
analysis could be  extended  to a general case initiated in
\cite{73}, where the elliptic curves are  replaced by $K3$
surfaces. This  might   be applied to the resolution of orbifold
singularities  in the moduli space of certain models, describing a
$D2$ brane
 wrapped $n$ times over the fiber of an elliptic $K3$, as  follows \cite{74}

\be
{\mathcal{M}}_{1,n}=Sym(K3)=\frac{K3^{\otimes n}}{S_n}. \ee
 Here  ${\mathcal{M}}_{1,n}$ denotes the moduli space of a D2-brane
with charges $(1,n)$ and $S_n$ is the group of permutation of $n$
elements.

\section{Conclusion}

In this paper we have studied the NC version of Calabi-Yau \
hypersurface orbifolds using the algebraic geometry approach of
[40,41] combined with toric geometry method of complex manifolds.
Actually this study extends the analysis on the NC Calabi-Yau
manifolds with discrete torsion initiated in [41] and expose
explicitly the solving of non commutativity in terms of toric
geometry data. Our main results may be summarized as follows:
\par $(1)$ First we have developed a method of getting $d$ complex
 Calabi-Yau
mirror coset manifolds $C^{k+1}/C^{\ast r}$ , $k-r=d,$ as hypersurfaces in $%
WP^{d+1}$. The key idea is to solve the $y_{i}$ inavriants
(2.12-13) of mirror geometry in terms of invariants of the
$C^{\ast }$ action of the weighted projective space and the toric
data of $C^{k+1}/C^{\ast r}$. As a matter of facts, the above
mentioned mirror Calabi-Yau spaces are described by homogeneous
polynomials $P_{\Delta }(u)$ of degree $D=\sum_{\mu
=1}^{d+2}\delta _{\mu }=\sum_{\mu
=1}^{d+2}\sum_{a=1}^{r}p_{a}q_{\mu }^{a},$
where $\delta _{\mu }$\ are projective weights of the $C^{\ast }$ action, $%
q_{\mu }^{a}$ entries of the well known Mori vectors and the
$p_{a}$'s are given integers. Then we have determined the general
group $\Gamma $ of discrete isometries of $P_{\Delta }(u).$ We
have shown by explicit computation that in general one should
distinguish two cases $\Gamma _{0}$ and $\Gamma _{cd}.$ First
$\Gamma _{0}$\ is the group of isometries of the
hypersurface $\sum_{\mu =1}^{d+2}u_{\mu }^{\frac{D}{\delta _{\mu }}%
}+a_{0}\prod_{\mu =1}^{d+2}\left( u_{\mu }\right) =0,$ generated
by the changes $u_{\mu }^{\prime }=u_{\mu }$ $\left( {\cal W}
\right) ^{{\bf b}_{\mu }}$\ where the weight vector ${\bf b}_{\mu
}$\ is given by the sum of $Q_{\mu }$\ \ and $\xi _{\mu }$
respectively associated with the Calabi-Yau charges and the
vertices data of the toric manifold $M_{\triangle }^{d+1}.$ In
case  where the complex deformations are taken into account (see
eq(2.27)), the symmetry
group reduces to the subgroup $\Gamma _{cd}$\ generated by the changes $%
u_{\mu }^{\prime }=u_{\mu }$ $\left( {\cal W} \right) ^{{\bf b}_{\mu }}$ where now $%
{\bf b}_{\mu }$\ has no $\xi _{\mu }$\ factor.

$(2)$ Using the above results and the algebraic geometry approach,
we have developed a method of building NC Calabi-Yau orbifolds in
toric manifolds. Non commutativity is solved in terms of the
discrete torsion and bilinears of the weight vector ${\bf
a}^{\nu}_{\mu }$; see eq(3.11). Among our results, we have worked
out several matrix representations of  the NC  quintic algebra
obtained in \cite{41} by using various Calabi-Yau toric geometry
data. We have also given the generalization of these results to
higher dimensional Calabi-Yau hypersurface orbifolds and derived
the explicit form of the non commutative
 $%
D$-tic orbifolds.\\ $(3)$ We have   extended to  higher  complex
dimension
Calabi-Yau's realized as toric orbifold of type $\frac{T^{4k+2}}{%
\Gamma }$ with discrete torsion. Due to constraint eqs on non
commutativity, we have  shown that in the elliptic realization of
the two torii factors, $\Gamma $ is constrained to be equal
 to $\mathbf{{Z_{2}^{2k}%
}}$, the real dimension should be $2k+2$ and non commutativity is
solved in terms of  the $2k$ dimensional Clifford algebra.  We
have  also discussed  the fractional brane which correspond to
reducible representations of  toric Calabi-Yau algebras.
\\
\\
 {\bf
Acknowledgments}\\ One of us (AB)  is very  grateful to  J. McKay
and A. Sebbar for discussion,  encouragement, and  scientific
helps. He would like also to thank J.J. Manjar\'\i n and  P. Resco
for earlier collaboration on this subject.  This work is partially
supported by PARS, programme de soutien \`a la recherche
scientifique; Universit\'e Mohammed V-Agdal, Rabat.

\end{document}